\documentclass[%
 reprint,
superscriptaddress,
 amsmath,amssymb,
 aps,
]{revtex4-1}

\usepackage{graphicx}
\usepackage{subfigure}
\usepackage{dcolumn}
\usepackage{bm}
\usepackage{color}

\def\beq{\begin{equation}}
\def\eeq{\end{equation}}

\begin{document}

\preprint{APS/123-QED}

\title{Fundamental structural characteristics of planar granular assemblies:\\ 
self-organisation and scaling away friction and initial state}

\author{Takashi Matsushima}
 \email{tmatsu@kz.tsukuba.ac.jp}
\affiliation{%
 Department of Engineering Mechanics and Energy, University of Tsukuba, Tsukuba, Japan
}%


\author{Raphael Blumenfeld}%
 \email{rbb11@cam.ac.uk}
\affiliation{
Imperial College London, London SW7 2AZ, UK
}%
\affiliation{
National University of Defense Technology, Changsha 410073, Hunan, China
}%
\affiliation{
Cavendish Laboratory, Cambridge University, JJ Thomson Avenue, Cambridge CB3 0HE, UK
}%


\date{\today}

\begin{abstract}
The micro-structural organisation of a granular system is the one most important determinant of its macroscopic behaviour. Here we identify the fundamental factors that determine the statistics of such micro-structures, using numerical experiments to gain general understanding. The experiments consist of preparing and compacting isotropically two-dimensional granular assemblies of polydisperse frictional discs and analysis of the emergent statistical properties of quadrons - the basic structural elements of granular solids. 
The focus on quadrons is because the statistics of their volumes have been found to display intriguing universal-like features \cite{MaBl14}.
The dependence of the structures and of the packing fraction on the inter-granuar friction and initial state are analysed and a number of significant results are found.
(i) An analytical formula is derived for the mean quadron volume in terms of three macroscopic quantities: the mean coordination number, the packing fraction and the rattlers fraction. 
(ii) We derive a unique, initial-state-independent, relation between the mean coordination number and the rattler-free packing fraction. The relation is supported numerically for a range of different systems. 
(iii) We collapse the quadron volume distributions from all systems onto one curve and verify that they all have an exponential tail.
(iv) The nature of the quadron volumes distribution is investigated by decomposition into conditional distributions of volumes given cell order and we find that each of these also collapses onto a single curve.
(v) We find that the mean quadron volume decreases with increasing inter-granular friction coefficients, an effect that is prominent in high order cells. We argue that this phenomenon is due to an increased probability of stable irregularly-shaped cells and test this by a herewith developed free cell analytical model.  
We conclude that, in principle, the micro-structural characteristics are governed mainly by the packing procedure, while effects of inter-granular friction and initial states are details that can be scaled away. However, mechanical stability constraints suppress slightly occurrence of small quadron volumes in cells of order $\geq 6$  and the magnitude of this effect does depend on friction. We quantify in detail this dependence and the deviation it effects from exact collapse for these cells.
(vi) We argue that our results support strongly the view that ensemble granular statistical mechanics does not satisfy the uniform measure assumption of conventional statistical mechanics.
Results (i)-(iv) have been reported in \cite{MaBl14} and are reviewed and elaborated on here. The results in (v) and the argument (vi) are new.


\end{abstract}

\pacs{Valid PACS appear here}
\maketitle


\section{\label{sec:level1}Introduction}
This paper discusses the structural characteristics of random packings of planar (2D) granular solids. The science of random granular packing has a very long history and statistical methods have been introduced into this field in the 60's \cite{Bernal1960, Mogami1965}. In recent years, however, a new way to describe random structures locally has emerged, which proved useful for a range of fundamental issues \cite{BaBl02,BlEd03,BlEd06}. The method consists of constructing space-tessellating volume elements, called quadrons, the structure of each of which is described by a local tensor. 
Unlike traditional tessellations, e.g. Voronoi-based, the quadron description preserves the connectivity information, since their construction is based on the force-carrying inter-granular contacts.
This description has proved useful for deriving the constitutive equation of the 2D isostaticity stress theory \cite{BaBl02,Bl04,Geetal08}, for the formulation of granular statistical mechanics (GSM) \cite{BlEd03,BlEd06,Bletal12,Bl08a,BlEd14}, and as a structural fingerprint of granular packs \cite{Fretal08,Fretal09,HiBl12}. 
The structural characteristics of mechanically equilibrated granular matter are significant because they affect directly a wide range of applications in soil mechanics \cite{Oda1982,Satake1993}, hydrology \cite{Vogel_Roth_2003} and chemical engineering \cite{Cheng_etal_1999}, to name a few. Significantly, these characteristics are not completely random and bear the hallmark of the self-organising process that brought the medium to mechanical equilibrium. The self-organisation leaves a fingerprint that manifests in the emergence of intriguing universality-like behaviour, as shown in 2D assemblies  \cite{MaBl14}. Thus, investigating the structural characteristics of granular matter is essential and is the focus of this paper.

It has been shown by Frenkel et al. \cite{Fretal08} that the structure of the quadron volume probability density function (PDF), $P(V_q)$, in 2D granular packs is best understood in terms of its conditional PDF's $P(V_q\mid e)$, where $e$ is the number of grains surrounding the cell associated with the quadron, also called cell order (see Fig. \ref{fig:quadron}). Based on geometrical considerations, and supported by numerical results, Frenkel et al. argued that the PDFs $P(V_q\mid e)$ should be independent of inter-granular friction and of the packing process (i.e. the history). The existence of history-independent features is significant in these systems, which are plagued by memory effects, and a detailed analysis of the conditional PDFs from first principles holds the promise of advancing the field.

Here, we first review and provide more details on our recent results \cite{MaBl14}. We investigate in detail the geometrical characterisation of the quadrons and the dependencies  of the PDFs $P(V_q)$ and $P(V_q\mid e)$ on the packing fraction, $\phi$, and the mean coordination (or contact) number per grain, $\bar{z}$.
This is done by numerical experiments on assemblies of polydisperse discs, isotropically packed by the same process to different values of $\phi$ and $\bar{z}$. We then go on to derive several new results, in particular, concerning the undermining effects of mechanical stability on perfect collapse of the conditional quadron volume distributions for some cell types in low friction systems. Specifically, We examine the decomposition proposed by Frenkel et al. \cite{Fretal08}, focusing on the significance of cell shapes, and we show that the universality suggested by them must be augmented with a mechanics-based consideration. This we support by constructing a simple free cell model, whose predictions agree with the numerical experiments.
Finally, we argue that the observed collapses may be a direct evidence for the failure of the assumption of `uniform measure' in GSM - a failure that has been observed previously in other systems \cite{PaFr12, Bletal15, Pa15, Fretal13, Bietal15}. In the concluding section we discuss all the results and the new insight that they provide into structures of general random granular packs in mechanical equilibrium.

\section{The relation between quadrons and statistical mechanics}

\begin{figure}[!bht]
\begin{minipage}[t]{0.4\textwidth}
\includegraphics[width=1.0\textwidth]{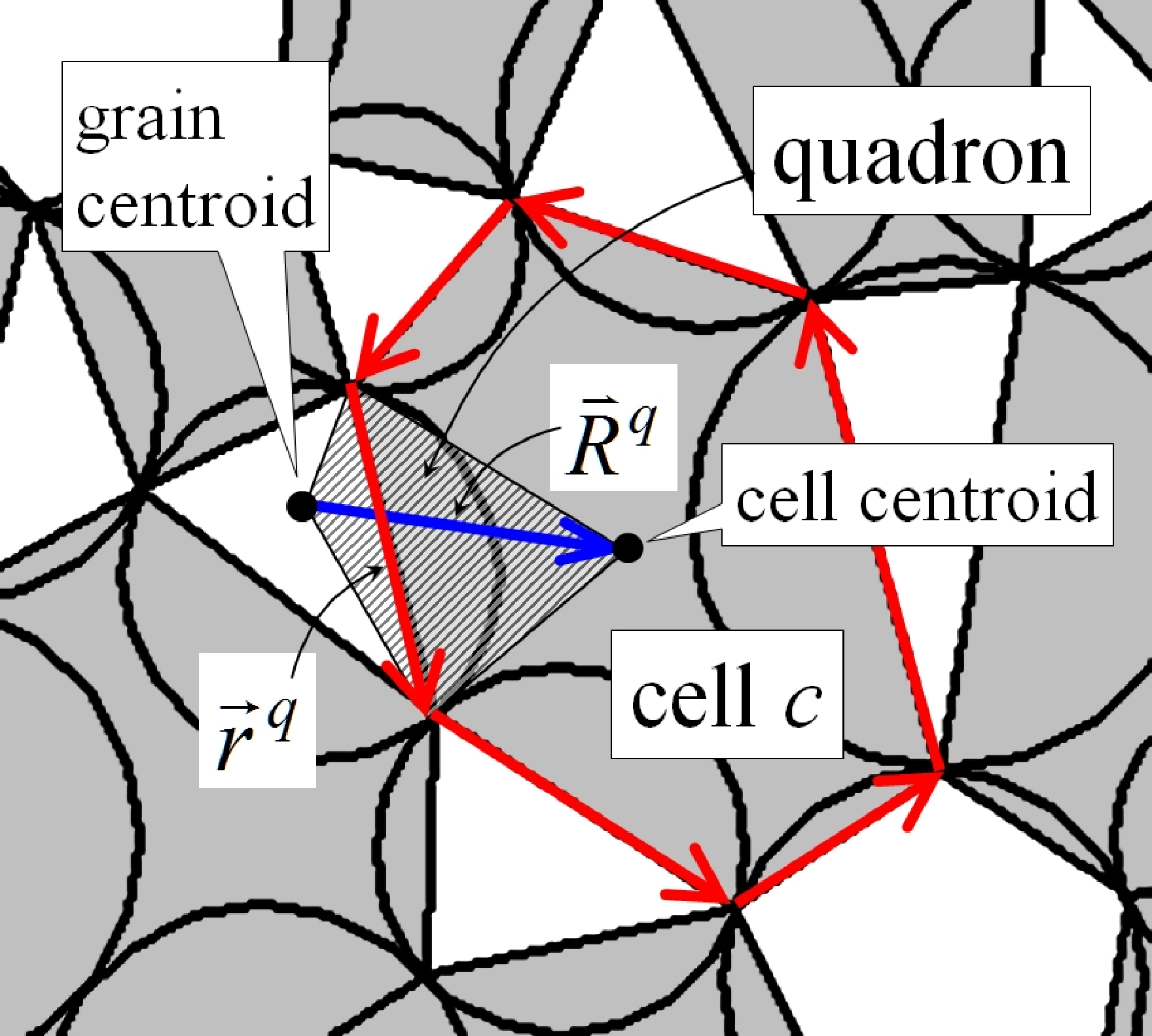}
\caption{Illustration of a quadrilateral quadron (shaded), whose diagonals are $\vec{R}^q$ (in blue) and $\vec{r}^q$ (in red). The inter-contact vectors $\vec{r}^q$ circulate anti-clockwise around grains.}
\label{fig:quadron}
\end{minipage}
\end{figure}

There are direct relations between the quadrons, their volumes, and the GSM. To be self-contained, we review briefly this relation as a basis for some of the analysis to follow. GSM of assemblies in static equilibrium is based on entropy, $S$, defined as the logarithm of the number of microstates that a collection of $N$ grains can be organised into. These microstates are of two types: structure- and stress-based \cite{BlEd05,BlEd06,Heetal07,Bl08a}.
A structural microstate consists of a specific configuration of the collection of grains. A stress microstate corresponds to a specific distribution of forces (or stress) that develops inside a given structural configuration under a specific set of boundary forces. It has been shown recently \cite{Bletal12} that the structure and stress sub-ensembles are inter-dependent. 
The formulation of GSM has evolved since its conception in 1989 \cite{EdOa89a} and it is generally accepted now that a significant contribution to the structural entropy comes from the connectivity and, therefore, that the structural degrees of freedom are determined by the contact positions. It has been then argued \cite{BlEd03,BlEd06} that the most straightforward way to parameterise these is by their differences, which correspond to the inter-contact vectors, $\vec{r}$, circulating every grain, as shown in Fig. \ref{fig:quadron}.

The interganular forces can be determined in terms of the external forces on the boundary, $\vec{f}_m$ ($m=1,2,...,M$), whether the structure is statically determinate (isostatic) or not \cite{Ma71}. For assemblies of $N$ grains, the number of boundary forces is a finite fraction of the number of boundary grains in $d$ dimensions, $N^{(d-1)/d}$. Thus, the boundary forces are the degrees of freedom spanning the stress microstates. 
The partition function, $Z$, describing the combined ensemble is

\begin{equation}
Z = \int e^{-\frac{\mathcal{C}}{\tau} - \sum_{\alpha \beta} \frac{\mathcal{F}_{\alpha \beta}}{X_{\alpha \beta}}} G\left(\left\{\vec{r}\right\}\right)\Theta \prod_{n=1}^{N_s} d\vec{r}_n \prod_{m=1}^{M} d\vec{f}_m \ ,
\label{Z1}
\end{equation}
where $\Theta$ is a product of $\delta$- and step-functions that specify the constraints on the ensemble, e.g. that the mean number of contact per grain, $\bar{z}$, is fixed, the system is in mechanical equilibrium, and that all systems were generated by the same process. 
The function $\mathcal{C}$ quantifies the connectivity (replacing the originally proposed and misbehaving volume function \cite{Bletal16}) and is expressed in terms of $N_s=N\bar{z}/2$ independent $\vec{r}$ vectors, $\tau$ is the contactivity - an analogue of the temperature, defined as $\partial\left(\langle\mathcal{C}\rangle\right)/\partial S$, $\vec{f}_m$ is one of the boundary forces on the system, $X_{\alpha \beta}$ is a component of the angoricity tensor - a measure of the fluctuations in the stress configurations, defined as $\partial\left(\langle\mathcal{F}_{\alpha\beta}\rangle\right)/\partial S$ \cite{BlEd05,Heetal07,BlEd09}, and $\mathcal{F}$ is the force moment tensor, defined as

\begin{equation}
\mathcal{F}_{\alpha \beta} = \sum_{g, g'} F_\alpha^{g g'} \mathcal{R}_\beta ^{g g'} \ .
\label{F}
\end{equation}
The sum here is over all pairs of grains in contact, $g$ and $g'$, $\vec{F}^{g g'}$ is the force that grain $g'$ applies to grain $g$ at their contact point, located at $\vec{\mathcal{R}}^{g g'}$. 
The function $G\left(\left\{\vec{r}\right\}\right)$ specifies the probability of a particular configuration of the vectors $\left\{\vec{r}\right\}$ to occur independent of the Boltzmann-like factor. It depends on the dynamics of generation of the ensemble and it generalises the original assumption of `uniform measure', within which $G$ is constant \cite{EdOa89a}, in view of recent evidence that this assumption fails in a number of systems \cite{Fretal13,Bietal15}.

Since our concern here is only with structural features of granular packs, we will work in the limit of high angoricity, when the second term in the exponential of (\ref{Z1}) can be neglected. To be `high', the angoricity tensor components should be larger than the largest boundary force on any one particle, times the system size. The results obtained below are found to be independent of the boundary forces and are, therefore, general regardless of the regime we use. 
The structure partition function alone is then, $Z_s = \int e^{-\frac{\mathcal{C}}{\tau}} G\left(\left\{\vec{r}\right\}\right)\prod_{n=1}^{N_s} d\vec{r}_n$.

A key concept in GSM is the quadron. The quadrons are volume elements, determined by the inter-contact vectors, $\vec{r}^q$, that tessellate the volume of the system, where the index $q$ stands for the quadron, with which this vector is associated (see below). An example is shown in Fig. \ref{fig:quadron}. These elements were defined and described both in two dimensions (2D) \cite{BaBl02,BlEd03} and in three (3D) \cite{BlEd06}. Their construction in 2D is described briefly below. 
The number of quadrons is equal to (in 2D) or outnumbers (in 3D) the number degrees of freedom and in 3D \cite{BaBl02,BlEd03}. This led to proposing the quadrons as the quasi-particles of the system, rather than the grains, whose position variables are too few to span the phase space formed by the positions of the contact points \cite{BaBl02,BlEd03,Fretal08,BlEd14}. Moreover, each quadron is associated with exactly one $\vec{r}^q$ vector (see Fig. \ref{fig:quadron}), linking these volume elements directly to the GSM. It is this connection that makes studying quadrons and their structural properties important.

In 2D, quadrons are constructed as follows. (i) Define the centroids of grains $g$ and of cells $c$ as the mean position vectors of the contact points around them, respectively (Fig. \ref{Z1}). (ii) Connect the contact points around every grain to make polygons. (iii) Make the polygon edges into vectors $\vec{r}^{q}$, circulating the grain in the clockwise direction. (iv) Extend conjugate vectors, $\vec{R}^{q}$, from the grain centroids to the centroids of the cells surrounding them (Fig. \ref{Z1}). A quadron is the quadrilateral whose diagonals are $\vec{r}^{q}$ and $\vec{R}^{q}$. (v) Quantify the structure of every quadron by the following structure tensor

\begin{flalign}
C^q=\left(\epsilon\cdot\vec{r}^q\right) \otimes \vec{R}^q \quad ; \quad 
\epsilon \equiv \left(
  \begin{array}{cc}
    0  & 1  \\
    -1 & 0  \\
  \end{array}
\right) \ ,
\label{C}
\end{flalign}
which quantifies the structure of the disordered system locally at every point. The quadron volume is

\begin{eqnarray}
V_q = \frac{1}{2} | \vec{r}^q \times \vec{R}^q | =  \frac{1}{2} Tr\{C^q \}
\label{Vq}
\end{eqnarray}
The number of quadrons is $N\bar{z}$, which is exactly the number of degrees of freedom, as mentioned. A similar construction gives the equivalent volume elements in 3D, also called quadrons \cite{BlEd06,Bl08a}.

In the following, we investigate numerically the PDFs $P(V_q)$ and $P(V_q\mid e)$ \cite{Fretal08,HiBl12} and relate them to the distribution of cell structures. This is motivated by recent observations of an apparent universality of these PDFs \cite{MaBl14}.\\

\section{Numerical experiments}

Our numerical experiments were carried out using the Discrete Element Method (DEM) \cite{Cundall1979,Mat-Chang-2011}. 
The method consists of using an incremental time marching scheme, wherein the motions of 2D discs (grains) are each computed using Newton's second law. We postulate a repelling harmonic interaction potential, characterized by normal and tangential spring constants, $k_n$ and $k_s$, activated upon contact and overlap between discs. We set $k_s/k_n=1/4$ in this study, a value chosen for its common use in the literature \cite{Hakuno97, Calvetti08, Mat-Chang-2011}.

Our aim is to study polydisperse systems and we have chosen the distribution of the radii of the discs to be log-normal distribution due to its wide use in civil engineering and soil sciences \cite{Mitchell-Soga-2005},
\begin{eqnarray}
P (D) =\frac{1}{\sqrt{2 \pi} \sigma D} \exp \left( \frac{(\ln D - \ln D_0)^2}{2 \sigma^2}   \right)
\end{eqnarray}
where we make $D_0=1.$ the unit length and $\sigma=0.2$, as the size distribution is broad as it is  \cite{MaBl13}. These values give $\bar{D}=1.02$ and $D_{mode}=0.961$ for the mean grain size and mode, respectively.

The packing protocol of our systems is as follows. 
First, we construct three random packs in a double periodic domain, engineered to be on the verge of jamming. The packs consist of about $21400 \pm 1000$ discs and are made at packing fractions $\phi=0.76, 0.82$ and $0.84$. The small variation in the numbers of particles is due to the different densities of the three initial configurations.
These configurations are then used as initial states for the packing procedure; loose initial state (LIS), intermediated initial states (IIS), and dense initial state (DIS), respectively. 

Once an initial state is set, we assign all the particles the same friction coefficient, $\mu$, and apply to the system a slow isotropic stress $\sigma_c$ by changing the periodic length on both sides. 
To approximate behaviour of rigid particles, we limit the applied stress to a level that corresponds to an average overlap discs of at most $\delta=\sigma_c / k_n =10^{-5}$.
No gravitational force is applied and the compression continues until the fluctuations of both grain positions (per mean grain diameter) and inter-granular forces (per mean average contact force) are below very small thresholds - $10^{-9}$ and $10^{-6}$, respectively. This procedure is carried out for each initial state at five different values of the inter-granular friction coefficients: $\mu=0.01, 0.1, 0.2, 0.5$ and $10$, giving altogether 15 assemblies.

\begin{figure}[!bht]
\begin{minipage}[t]{0.45\textwidth}
\includegraphics[width=1.0\textwidth]{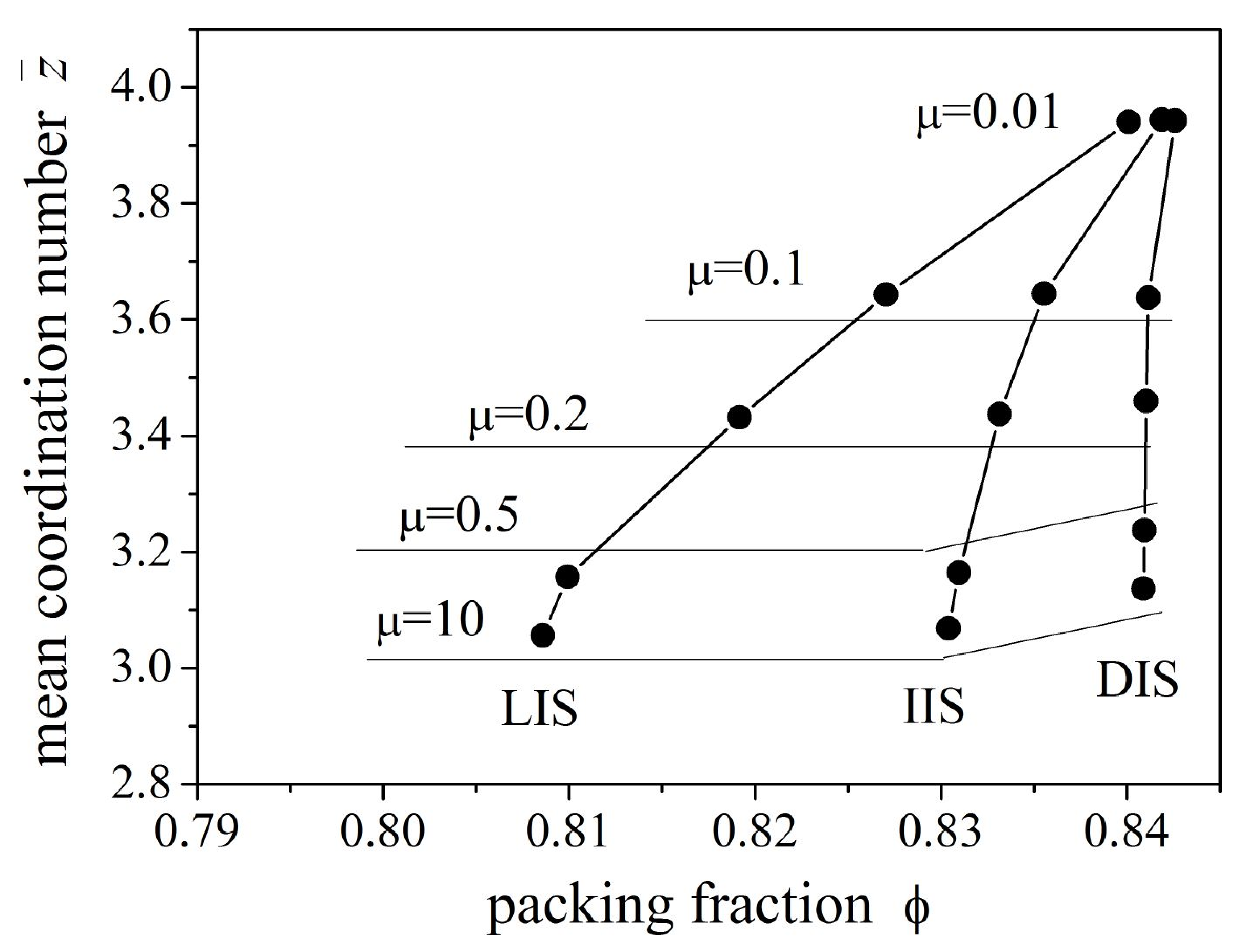}
\caption{Coordination number vs. packing fraction for 15 systems, generated from three different initial states, LIS, IIS, and DIS. Note the convergence for $\mu\to 0$.}
\label{fig:packing_fraction}
\end{minipage}
\end{figure}

Using this procedure, we have computed the packing fractions, the mean coordination numbers, and studied the structures in these systems. 
For the determination of $\bar{z}$, we disregard `rattlers', i.e. grains with one or no force-carrying contact. The results are shown in Fig. \ref{fig:packing_fraction}. 
The upper and lower bounds of the coordination number $\bar{z}_{max}=4$  and $\bar{z}_{min}=3$ correspond to the isostatic states for discs with friction coefficients of $\mu=0$ and $\infty$, respectively.
The three systems with $\mu=0.01$ converge into states with ($\bar{z}$, $\phi$)=(3.941, 0.840),(3.944, 0.842) and (3.943, 0.843), respectively, which are very close to the ideal frictionless jammed state. However, with increasing friction coefficient, the difference between the values of $\bar{z}$ and $\phi$ in the final states of systems started from different initial states increases (see Fig.  \ref{fig:packing_fraction}).

\begin{figure}[!bht]
\begin{minipage}[t]{0.3\textwidth}
\includegraphics[width=1.0\textwidth]{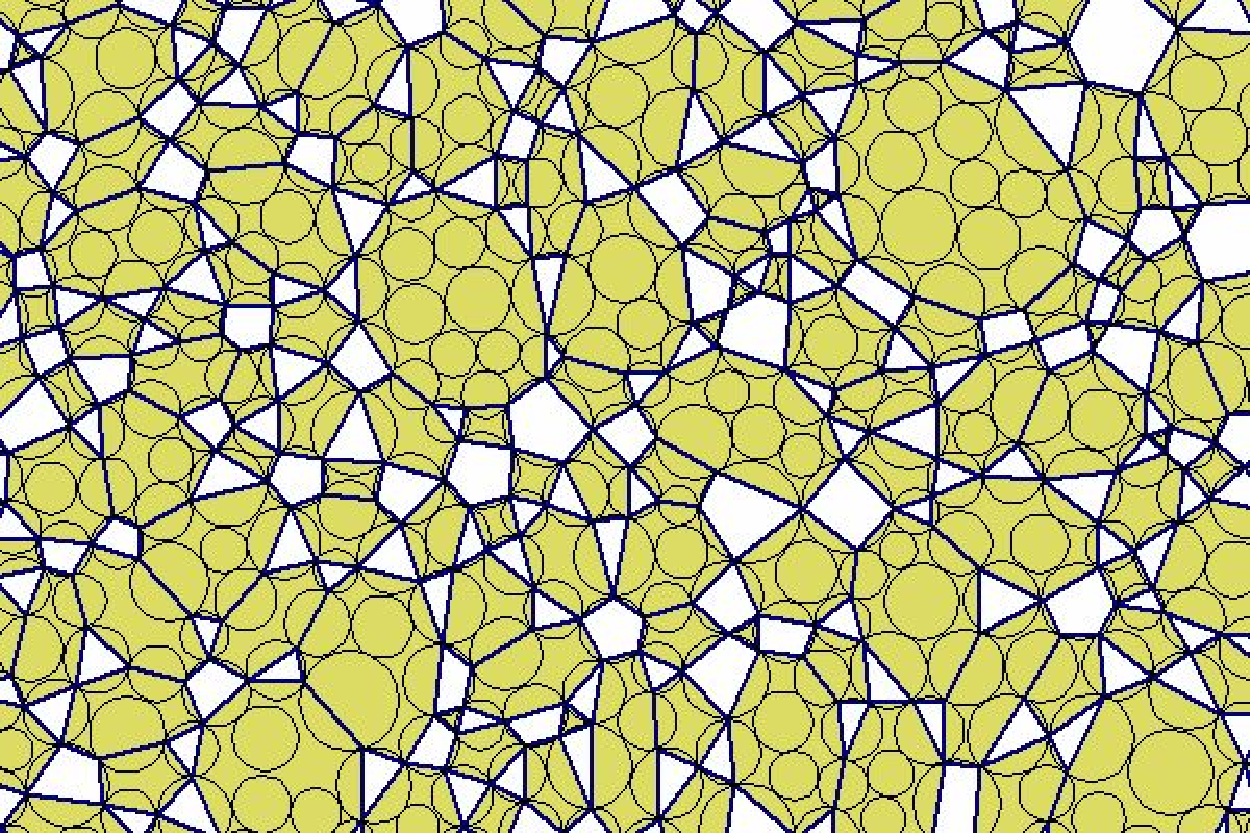}
\caption{Example of an assembly with $\mu=10$, generated from a LIS. Some of the cells contain rattlers, defined as discs that have at most one contact.}
\label{fig:cell_structure1}
\end{minipage}
\end{figure}

\begin{figure}[!bht]
\begin{minipage}[t]{0.3\textwidth}
\includegraphics[width=1.0\textwidth]{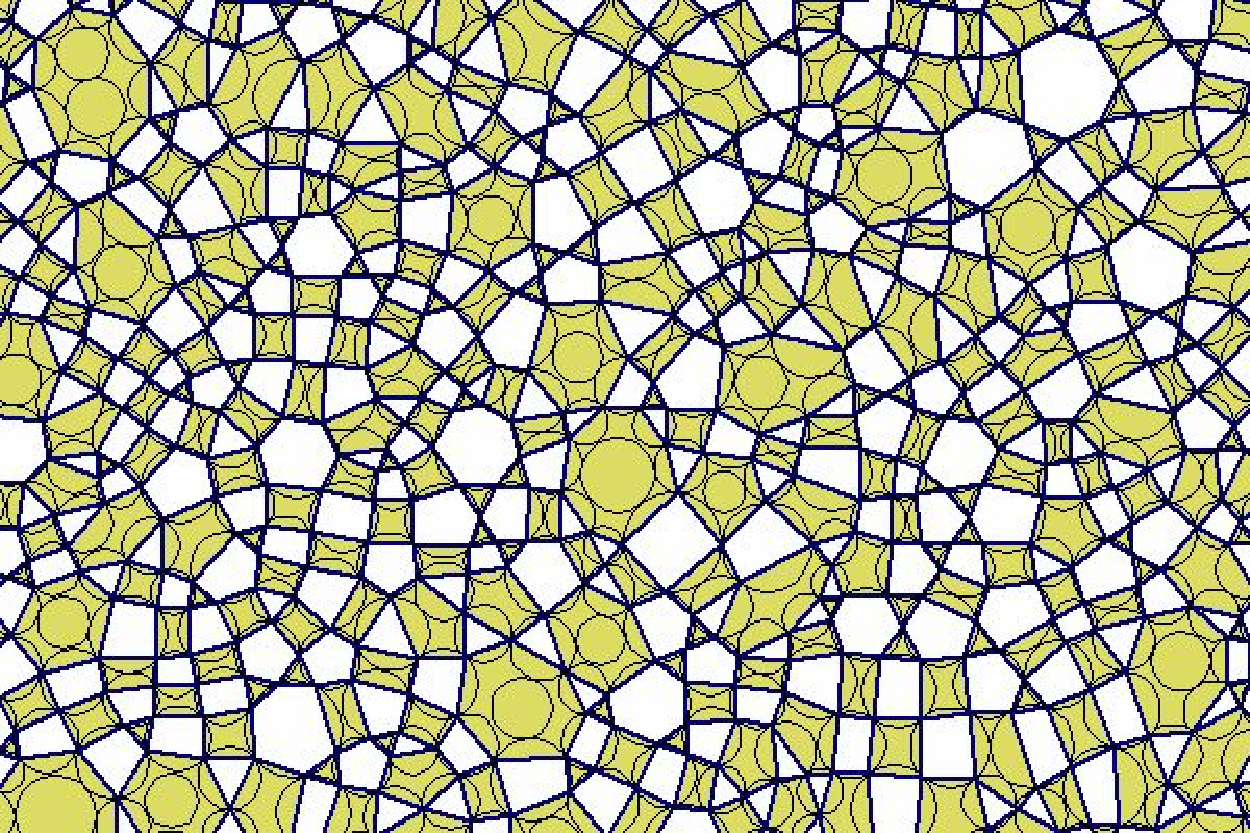}
\caption{Example of an assembly with $\mu=0.01$, generated from LIS.}
\label{fig:cell_structure2}
\end{minipage}
\end{figure}

\begin{figure}[!bht]
\begin{minipage}[t]{0.3\textwidth}
\includegraphics[width=1.0\textwidth]{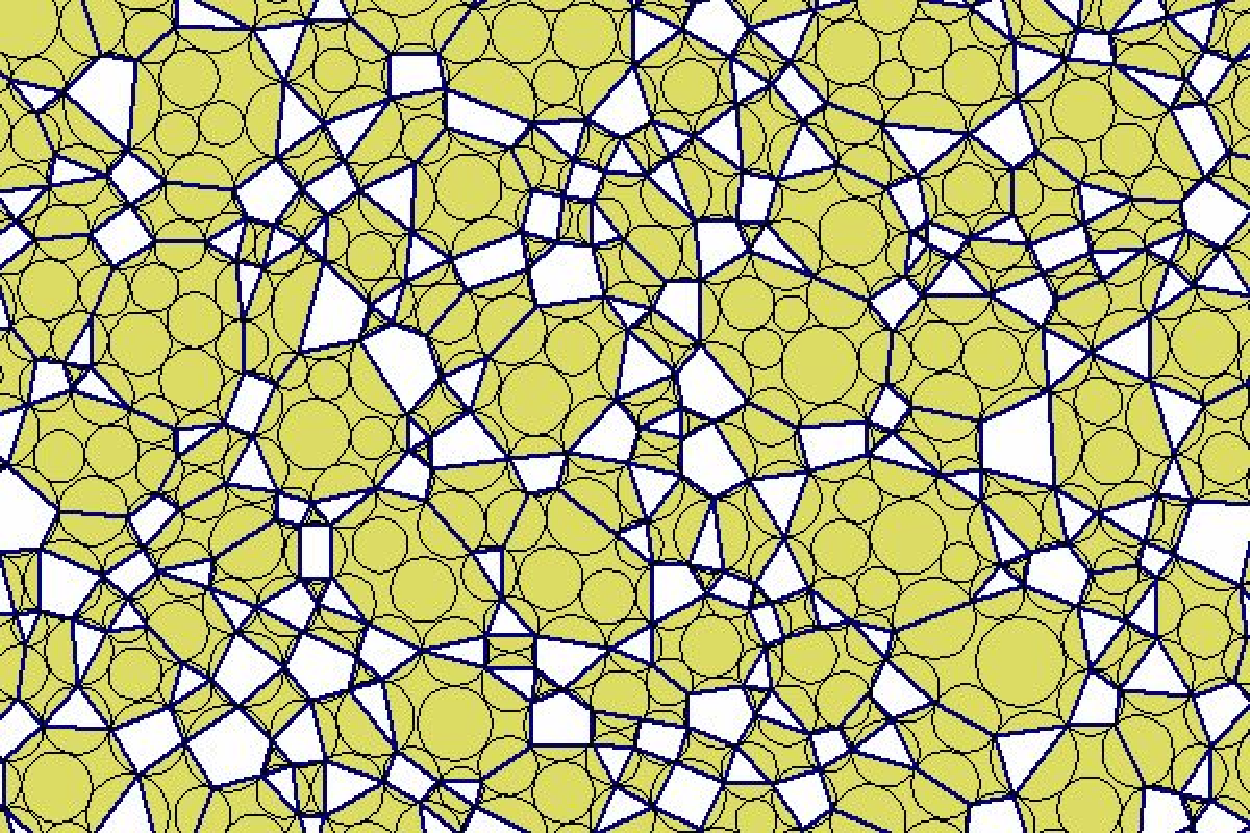}
\caption{Example of an assembly with $\mu=10$, generated from DIS.}
\label{fig:cell_structure3}
\end{minipage}
\end{figure}

Turning to the analysis of cell structures in these systems, consider the three examples shown in Figs. \ref{fig:cell_structure1} - \ref{fig:cell_structure3}, constructed from LIS, with $\mu=10$ (\ref{fig:cell_structure1}) and $0.01$ (\ref{fig:cell_structure2}), and from DIS with $\mu=10$ (\ref{fig:cell_structure3}). 
While traces of the initial condition can be detected in both the first two systems, for higher friction one clearly ends up with typically larger cells and a correspondingly lower $\bar{z}$.  
Note that the packing fraction in Figs. \ref{fig:cell_structure1} and \ref{fig:cell_structure3} are very similar (see also Fig. \ref{fig:packing_fraction}), but the cells in  \ref{fig:cell_structure1} are noticeably bigger typically. 
Also note the considerable number of rattlers in the large cells, an issue that will be discussed in detail below. 
To quantify the structural differences, we next study the quadron volume distribution. 

 
\section{Structural characteristics}

\subsection{\label{sec4-1} Quadron volume statistics}

\begin{figure}[!bht]
\begin{minipage}[t]{0.4\textwidth}
\includegraphics[width=1.0\textwidth]{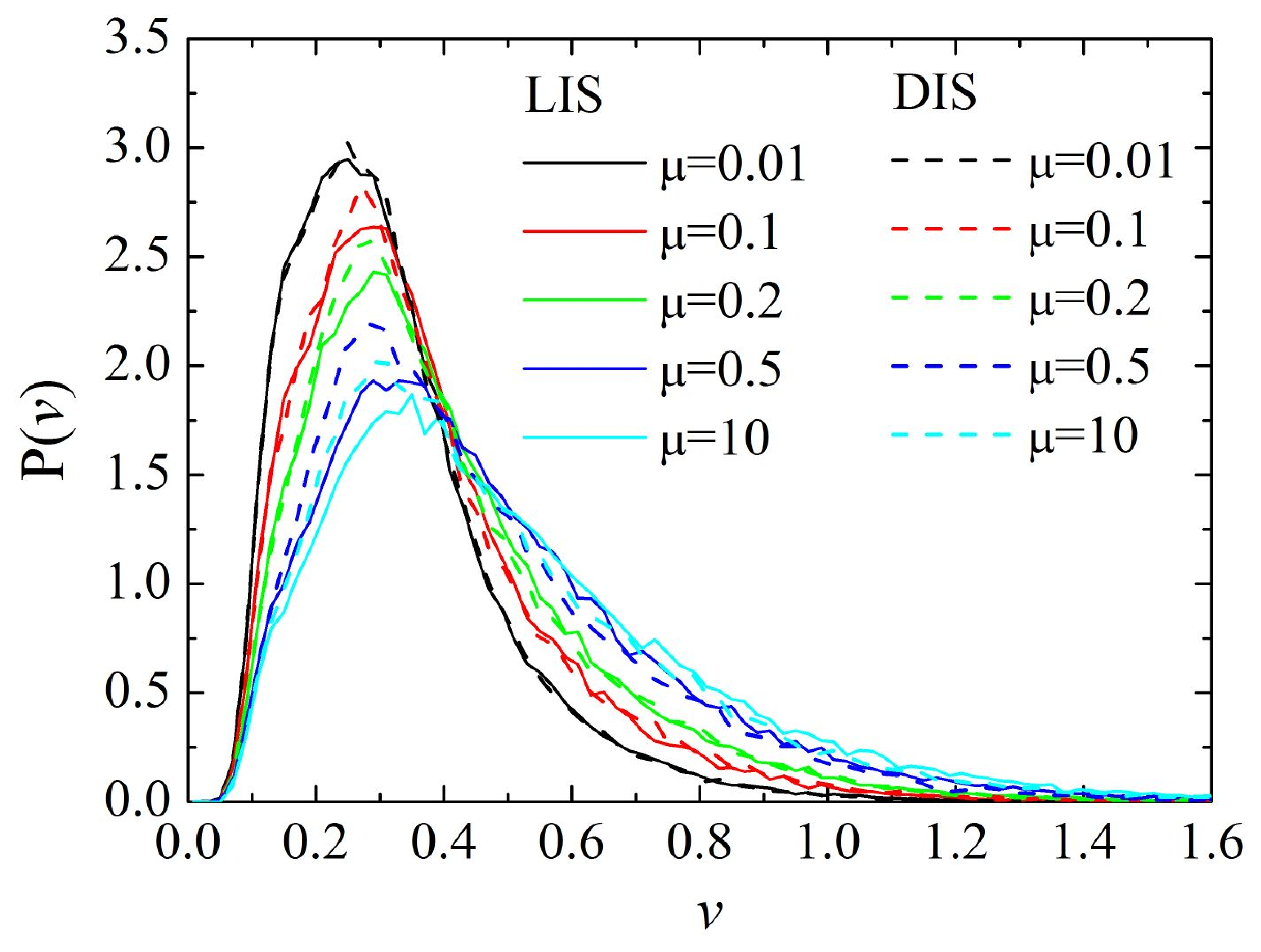}
\caption{PDF of the quadron volumes, normalized by the mean grain volume, $v=V_q/\bar{V}_g$ for all the 10 systems generated from LIS and DIS.}
\label{fig:PDF of quadron}
\end{minipage}
\end{figure}

\begin{figure}[!bht]
\begin{minipage}[t]{0.4\textwidth}
\includegraphics[width=1.0\textwidth]{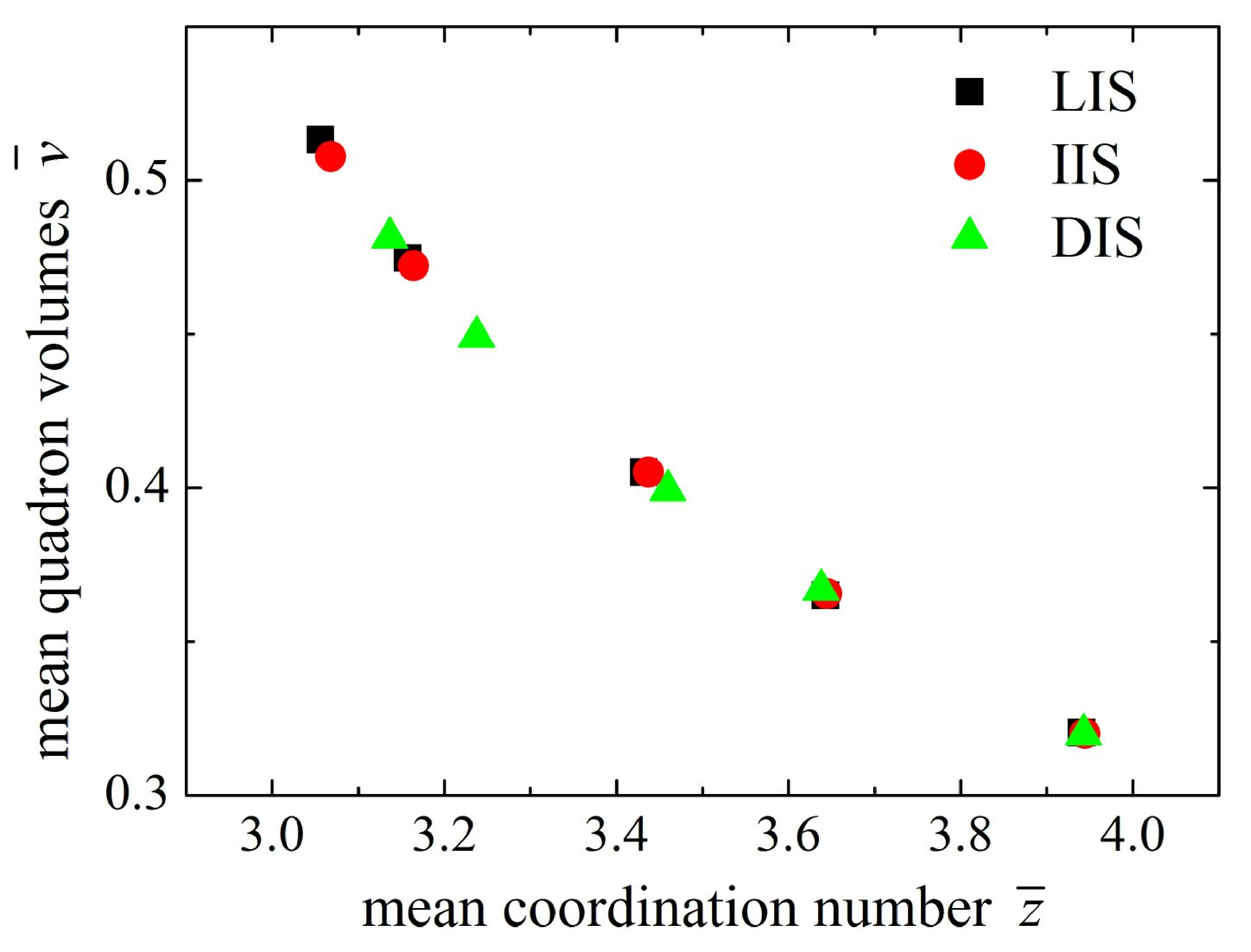}
\caption{Mean quadron volume $\bar{v}$ vs. mean coordination number $\bar{z}$ for all initial states.}
\label{fig:mean quadron volume vs zg}
\end{minipage}
\end{figure}

The grain size distribution is the same for all packs, making it convenient to normalize the quadron volumes by the mean grain volume $\bar{V}_g$, $v\equiv V_q/ \bar{V}_g$. The overall PDFs of quadron volumes are plotted in Fig. \ref{fig:PDF of quadron} for all friction coefficients and for LIS and DIS. 
The corresponding means are plotted against the mean coordination numbers, for all initial states, in Fig. \ref{fig:mean quadron volume vs zg}. All the points fall nicely on one curve {\it regardless of initial state}. 

\begin{figure}[!bht]
\begin{minipage}[t]{0.4\textwidth}
\includegraphics[width=1.0\textwidth]{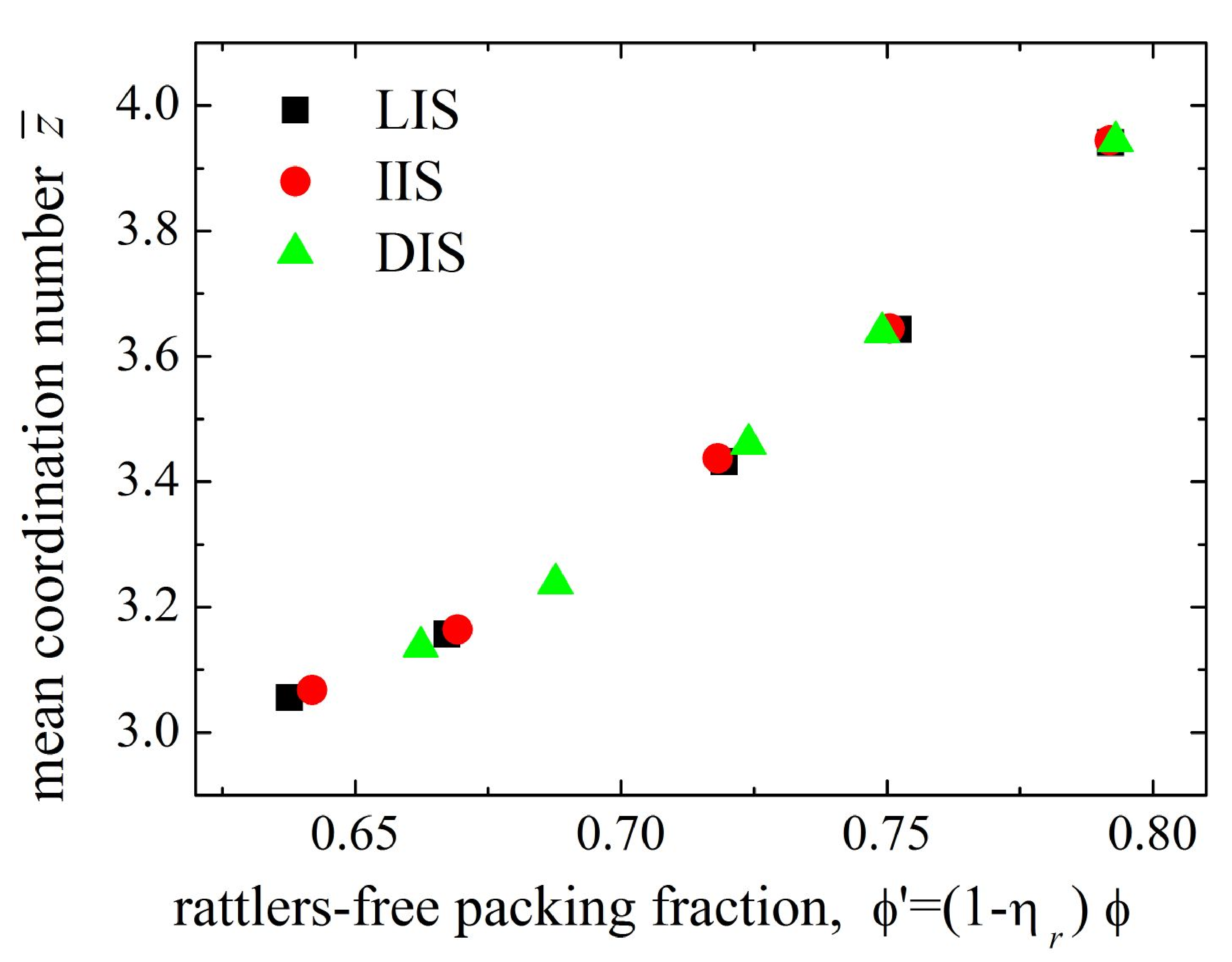}
\caption{Mean coordination number $\bar{z}$ vs. the rattlers-free packing fraction $\phi'$.}
\label{fig:mean zg vs. mod phi}
\end{minipage}
\end{figure}

On close inspection, this very weak dependence, if any, on initial state can also be observed in Fig. \ref{fig:PDF of quadron}. 
This implies that the dependence of the relation between $\bar{v}$ and $\bar{z}$ on the initial state, seen in Fig. \ref{fig:packing_fraction}, can be made to disappear under the right parameterization. A clue to such a parameterization is the fact that the quadron volumes are computed {\it ignoring the rattlers}. 
Following this idea, we can derive the relation between $\bar{v}$ and $\bar{z}$ as follows

\begin{eqnarray}
\phi'=\frac{V_s'}{V}=\frac{\sum^{N_g'} V_g}{\sum^{N_q} V_q}
=\frac{N_g' \bar{V}_g}{N_q \bar{V}_q}
=\frac{1}{\bar{v} \bar{z}}
\label{eq:model03}
\end{eqnarray}
\begin{eqnarray}
\bar{v} = \frac{1}{ \phi' \bar{z} }= \frac{1}{(1-\eta_{r}) \phi \bar{z}}
\label{eq:mod_phi}
\end{eqnarray}
where $\eta_{r}=N_{r}/N$ is the rattlers fraction and $\phi'$ is the packing fraction after removing all the rattlers. 
This elegant result suggests that there is indeed a unique relationship between $\bar{z}$ and the packing fraction, but only if the latter is corrected for the rattler fraction. 
This is verified by a direct plot of these two quantities, Fig. \ref{fig:mean zg vs. mod phi}.
It is also important to note that this relation is independent of initial state as Fig. \ref{fig:mean zg vs. mod phi} shows.

Comparing Figs. \ref{fig:mean zg vs. mod phi} and \ref{fig:packing_fraction}, we see that the rattler-free packing fractions are much smaller than the packing fractions commonly reported in the literature, e.g. \cite{silbert2010,abate2006approach}. 
The removal of rattlers also makes sense on mechanical grounds, since it does not affect the force transmission in our systems (nor in any system experiencing no body forces).

\begin{figure}[!bht]
\begin{minipage}[t]{0.4\textwidth}
\includegraphics[width=1.0\textwidth]{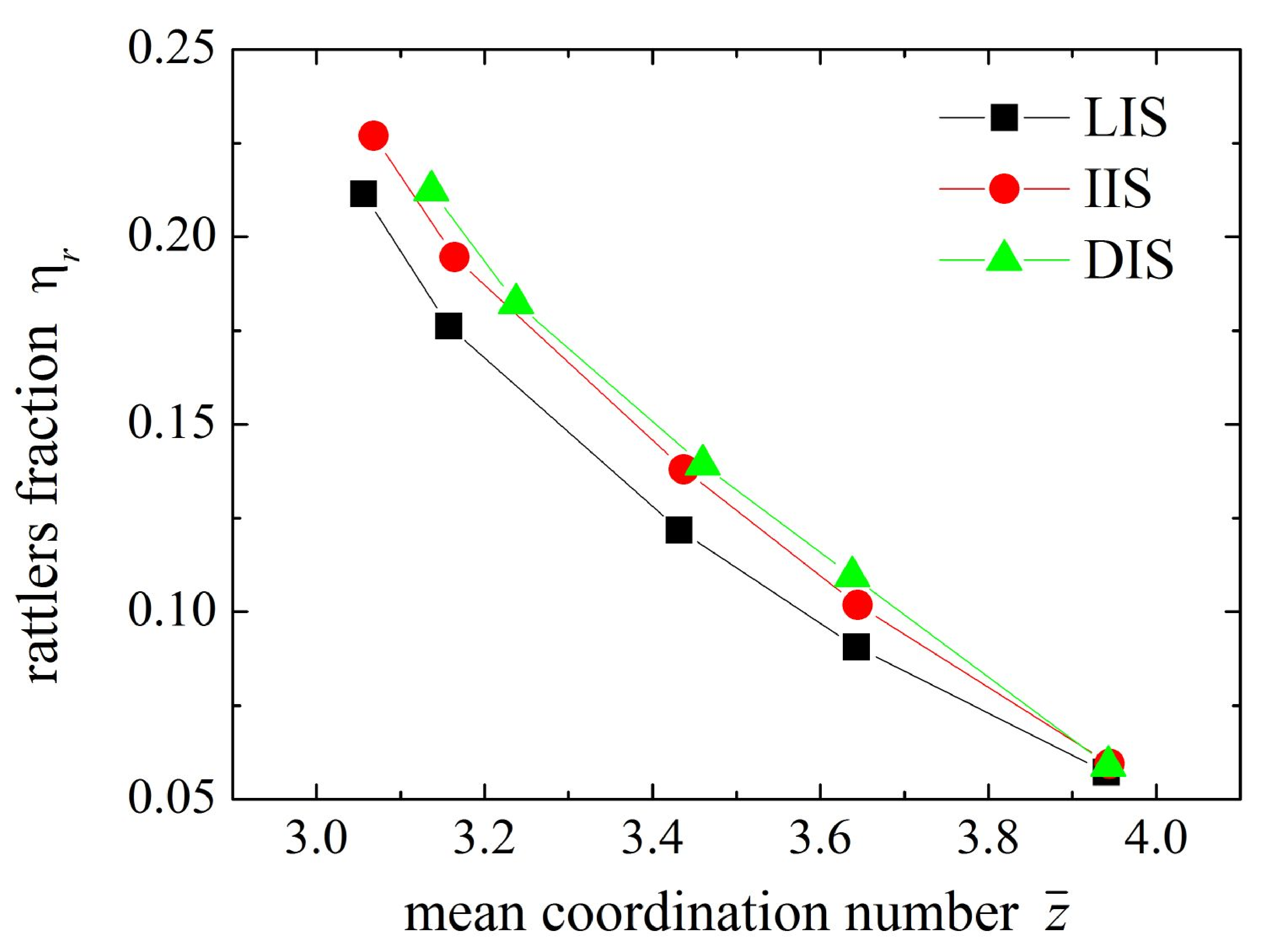}
\caption{ The rattlers fraction $\eta_{r}$ vs. the mean coordination number $\bar{z}$.}
\label{fig:r_rattlers vs. mean zg}
\end{minipage}
\end{figure}

We conclude that the seeming sensitivity of $\bar{z}(\phi)$ to the initial state (e.g. Fig. \ref{fig:packing_fraction}), commonly seen in the literature,  stems directly from the variation in $\eta_{r}$. Plotting $\eta_{r}$ as a function of $\bar{z}$ in our systems (Fig. \ref{fig:r_rattlers vs. mean zg}), we see that it is the small differences between the curves that gives rise to the observed differences in the conventional plots.

\begin{figure}[!bht]
\begin{minipage}[t]{0.4\textwidth}
\includegraphics[width=1.0\textwidth]{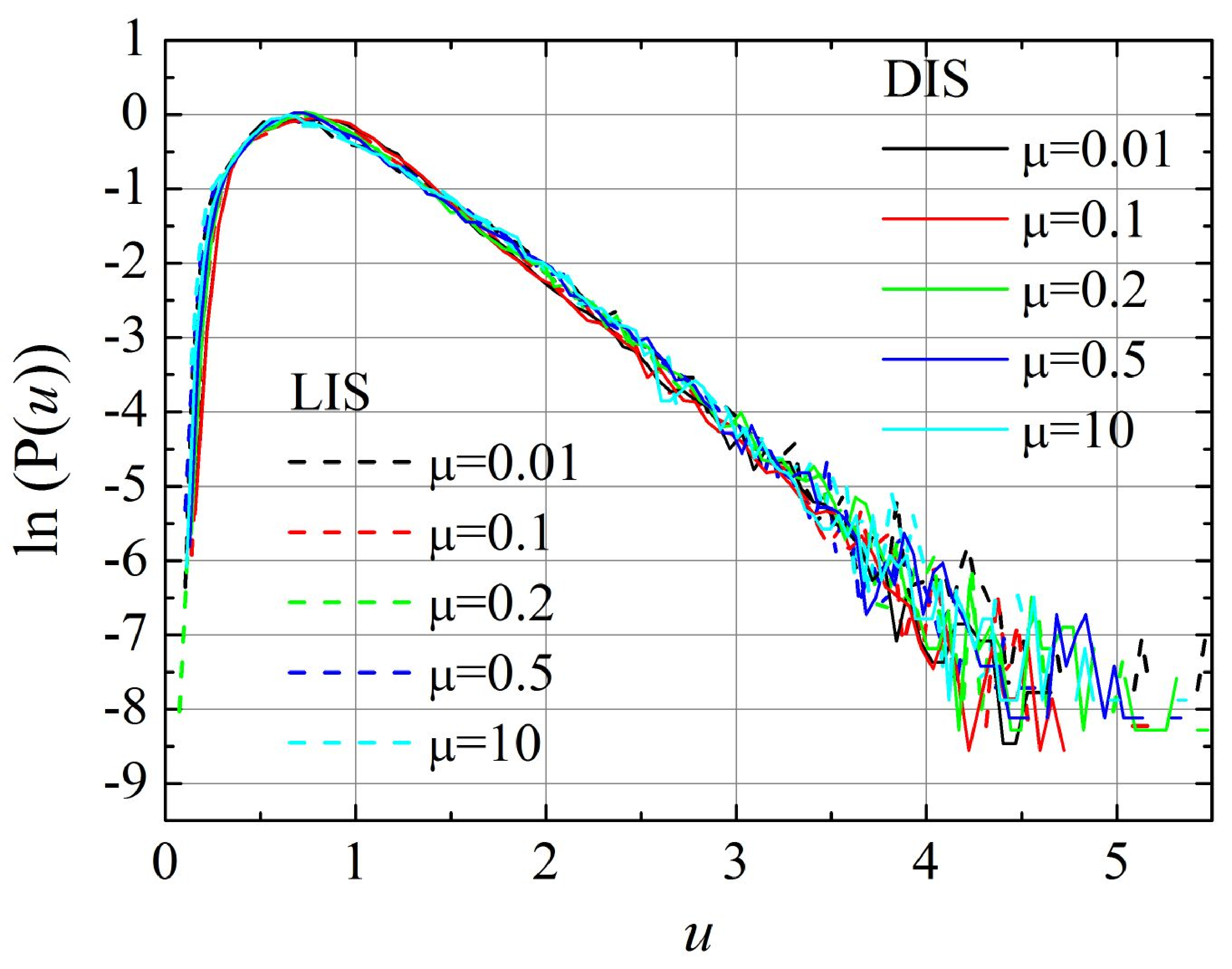}
\caption{The PDF of the normalized quadron volume $u=V_q/\bar{V}_q$ for  for all the 10 systems generated from LIS and DIS.}
\label{fig:vq_PDF_normalized_by_mean_vq}
\end{minipage}
\end{figure} 
It is important to comment here that this result is not really universal. It is correct for the specific packing protocol used here. While we believe that, for any given protocol, the plots of $\bar{z}(\phi')$ would collapse onto one curve, there is no reason that this curve should be universal.  In other words, we expect different protocols to display different $\bar{z}(\phi')$ curves, probably depending on the rate of rattlers generation. 
However, these results do provide a universal protocol-independent insight - the relation between the $\bar{z}$ and the packing fraction is directly linked to the mechanical stability of the structure and the way that forces are transmitted. We will discuss this insight more in the concluding section.  

Turning to consider the PDFs of the quadron volumes more closely, it is natural to expect the mean quadron volume to increase with $\mu$, simply because the cells get bigger. To get insight into the shapes of the PDFs, we scale the quadron volumes by their means, $u\equiv  V_q / \bar{V}_q=v/\bar{v}$. This simple scaling is sufficient to collapse all the PDFs almost perfectly onto one curve for all the systems, independent of friction and initial state (Fig. \ref{fig:vq_PDF_normalized_by_mean_vq}).  

Moreover, the collapsed PDF has an exponential tail, which  is a signature of the Boltzmann-like factor of equation (\ref{Z1}). 
Our best fit to the collapsed curve has the following form
\begin{eqnarray}
P(u) = f(u) e^{-\alpha u}
\label{CollapsedPu}
\end{eqnarray}
where $f(u)$ is a rational function. 
Using $u=v/\bar{v}=V_q/\bar{V}_q$ and the equipartition principle obtained in \cite{Bletal16}, $\langle V_{2D}\rangle = C_{2D} N_q\tau$, with $C_{2D}$ a constant of order unity and $\tau$ the contactivity (see eq. (\ref{Z1})), we have 
\begin{equation}
\frac{1}{ \phi' \bar{z} \tau }= \frac{C_{2D}}{\bar{V}_g}
\label{eq:EOS}
\end{equation}
where the right hand side is a constant that depends only on the grain size distribution. The exact relation between $\alpha$ and $C_{2D}$ remains to be found.

Thus, the quadron description makes it possible to collapse the statistics of all the systems, given the correct normalisation. As such, it gives better insight into the general characteristics of granular packs - characteristics that are independent of both the inter-granular friction coefficient and the initial state. 
In the next section, we explore the reasons for the apparent unified nature of the quadron volumes statistics.
 
\subsection{\label{sec4-2} Cell order statistics}

To understand better the structural characteristics, we use a recently-proposed decomposition of the quadron volumes into conditional distributions \cite{Fretal08}
\begin{eqnarray}
P\left(v\right)=\sum_e e Q(e) P\left(v\mid e\right)
\label{eq_decomposition}
\end{eqnarray}
Here $Q(e)$ is the occurrence probability of cells of order $e$, i.e. enclosed by $e$ grains and $P(v|e)$ is the conditional PDF of the normalized quadron volume, given that it belongs to a cell of order $e$.
$Q(e)$ is essential to the understanding of random granular packing \cite{BlToMaSoon} and it is this PDF that we wish to focus on next. 
As expected from Euler topological relation (below) and as can be observed from Figs. \ref{fig:cell_structure1}-\ref{fig:cell_structure3}, a lower value of $\bar{z}$ must be accompanied with a higher mean value of $\bar{e}$ and corresponds to a lower packing fraction $\phi'$. 

\begin{figure}[!bht]
\begin{minipage}[t]{0.4\textwidth}
\includegraphics[width=1.0\textwidth]{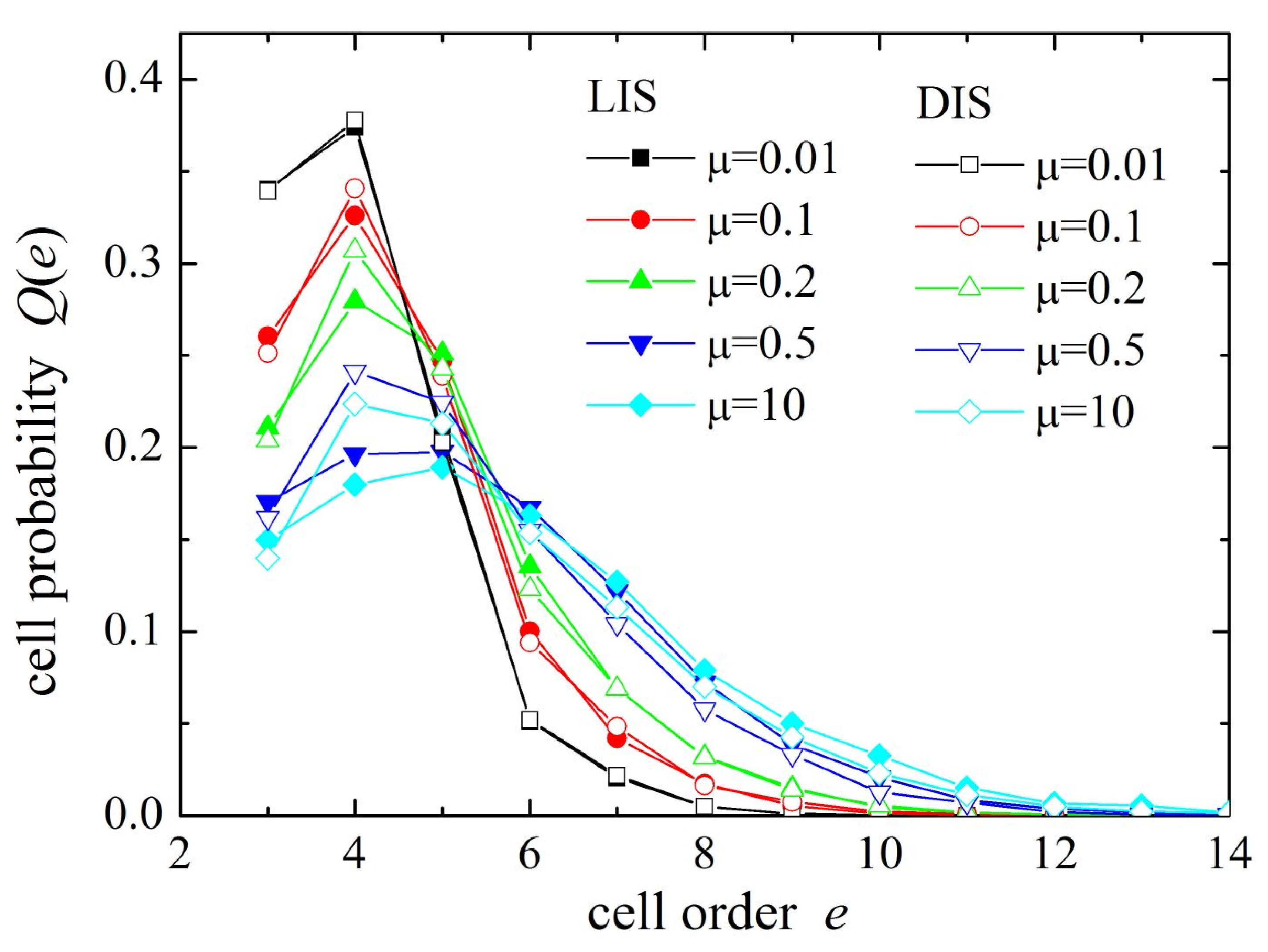}
\caption{The cell order probability $Q(e)$ for  for the 10 systems generated from LIS and DIS.}
\label{fig:q(e)_dense}
\end{minipage}
\end{figure} 
In Fig. \ref{fig:q(e)_dense} we plot $Q(e)$ for all the systems generated from LIS and DIS. The plots make evident two points. One is that the higher the inter-granular friction the larger the fraction of high-order cells. The other is  that $Q(e)$ is hardly dependent on the initial state for any $\mu$.

\begin{figure}[!bht]
\begin{minipage}[t]{0.4\textwidth}
\includegraphics[width=1.0\textwidth]{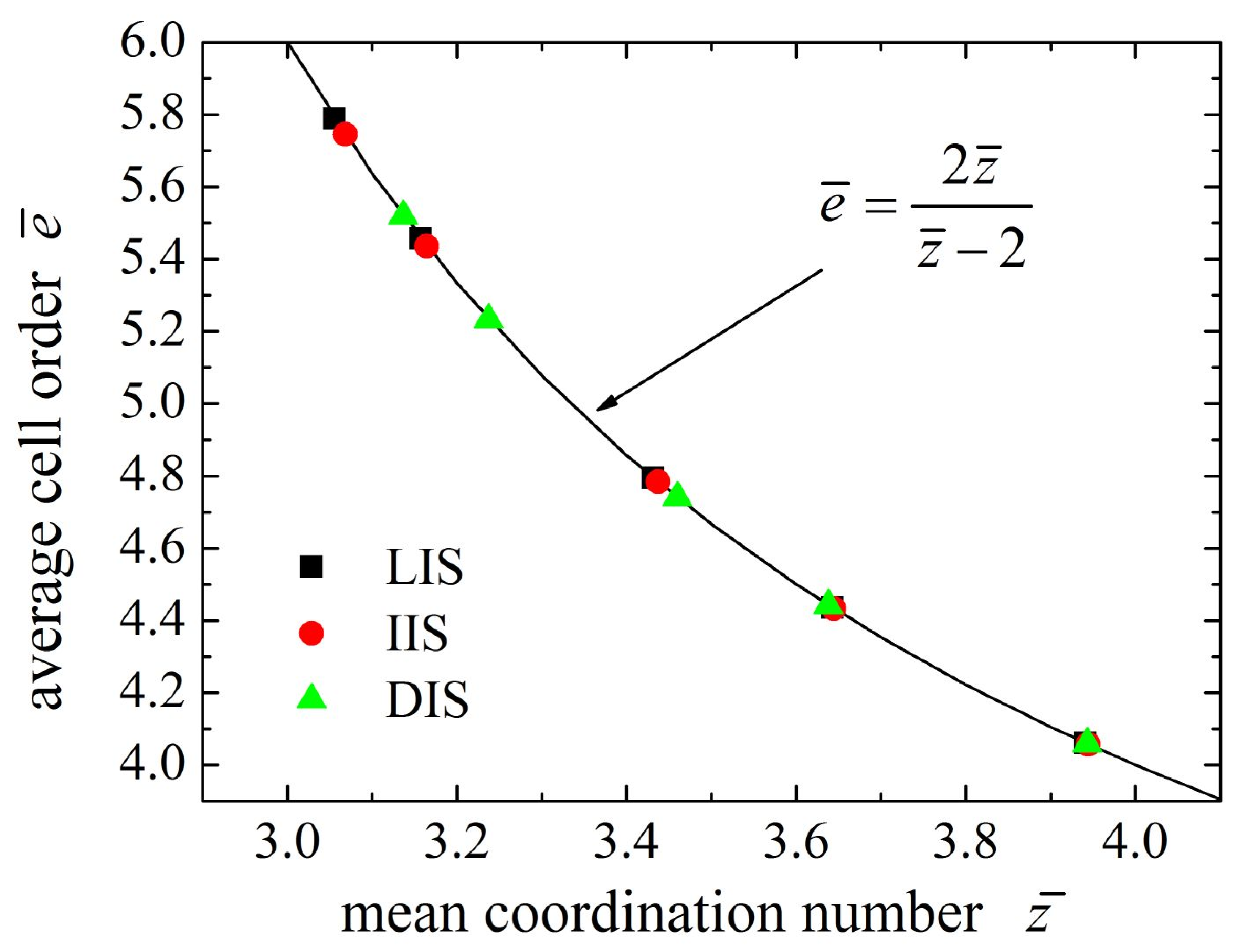}
\caption{The mean cell order $\bar{e}$ vs. the mean coordination number $\bar{z}$.}
\label{fig:Euler's relation}
\end{minipage}
\end{figure}

There is a direct relation between the mean cell order, $\bar{e}$, and the mean coordination number $\bar{z}$ and it can be derived from Euler's topological relation for a planar graph,
\begin{equation}
N_V - N_E +N_C = 1
\label{Euler}
\end{equation}
In this relation, $N_V$, $N_E$ and $N_C$ are, respectively, the numbers of the graph's vertices, edges and cells. In this relation we disregard the one large cell making the outside of the system, whose inclusion gives on the right hand side a topological characteristic 2, rather than 1. This topological characteristic is negligible for $N\gg 1$. 
Regarding the grain centres as vertices and the lines connecting centres of touching grains as edges, we have $N_V=N$, $N_E=N\bar{z}/2$, which gives $N_C=N\left(\bar{z}-2\right)/2$. Noting then that $N_C\bar{e}=2N_E$, we have \cite{Fretal08,MaBl14}
\begin{eqnarray}
\bar{e} = \frac{2 \bar{z}}{ \bar{z} -2} + O\left(\frac{1}{\sqrt{N}}\right) \ ,
\label{eq:mean_e}
\end{eqnarray}
The rightmost term is a negligible boundary correction. Fig. \ref{fig:Euler's relation} shows that the numerical systems satisfy this relation very well, showing that the sizes of the numerical pack are sufficiently large to ignore finite size effects.

$Q(e)$ is sensitive to the value of $\bar{z}$, which is a function of $\mu$ (see Fig. \ref{fig:q(e)_dense}), while $P\left(v\mid e\right)$ has been shown by Frenkel et al \cite{Fretal08}  (see also below) to be hardly dependent on $\mu$. This suggests that there may be a parameterization of $Q(e)$ that collapses all the curves corresponding to the collapse of $P\left(u^q \right)$ in Fig. \ref{fig:vq_PDF_normalized_by_mean_vq}.

We find that all the curves collapse if we plot $\bar{e} e Q(e)$ as a function of $e'=(e-\bar{e})/\bar{e}^2$. This collapse is better in these variables and this particular normalisation than the one presented in \cite{MaBl14}. 
The fundamental reason for this is not fully understood, but we note that $\bar{e} Q(e)$ is the PDF of the variable $y=e/\bar{e}$ and, therefore, that $\bar{e} e Q(e) dy$ is the probability of finding a quadron belonging to an $e$ cell out of the entire quadrons population. We also note that $e'=(e-\bar{e})/\bar{e}$ is the relative deviation of $y$ from its mean value, $y=1$. 
The collapsed master curve appears to be fitted reasonably well by a truncated Gaussian
\begin{eqnarray}
\bar{e} e Q(e) = 
\frac{\sqrt{2/\pi}}{\sigma_r (1+erf(e'_{min}/\sqrt{2}\sigma_r))} 
exp\left(-e'^2/2\sigma_r^2\right) \ ,
\label{eq:gaussian}
\end{eqnarray}
with $e'_{min}=0.076$, $\sigma_r=0.082$ and $erf(x)$ being the error function, except for a small deviation at the large $e$-tail. We have no explanation for either the Gaussian form or the deviations from it at large $e$.

\begin{figure}[!bht]
\begin{minipage}[t]{0.4\textwidth}
\includegraphics[width=1.0\textwidth]{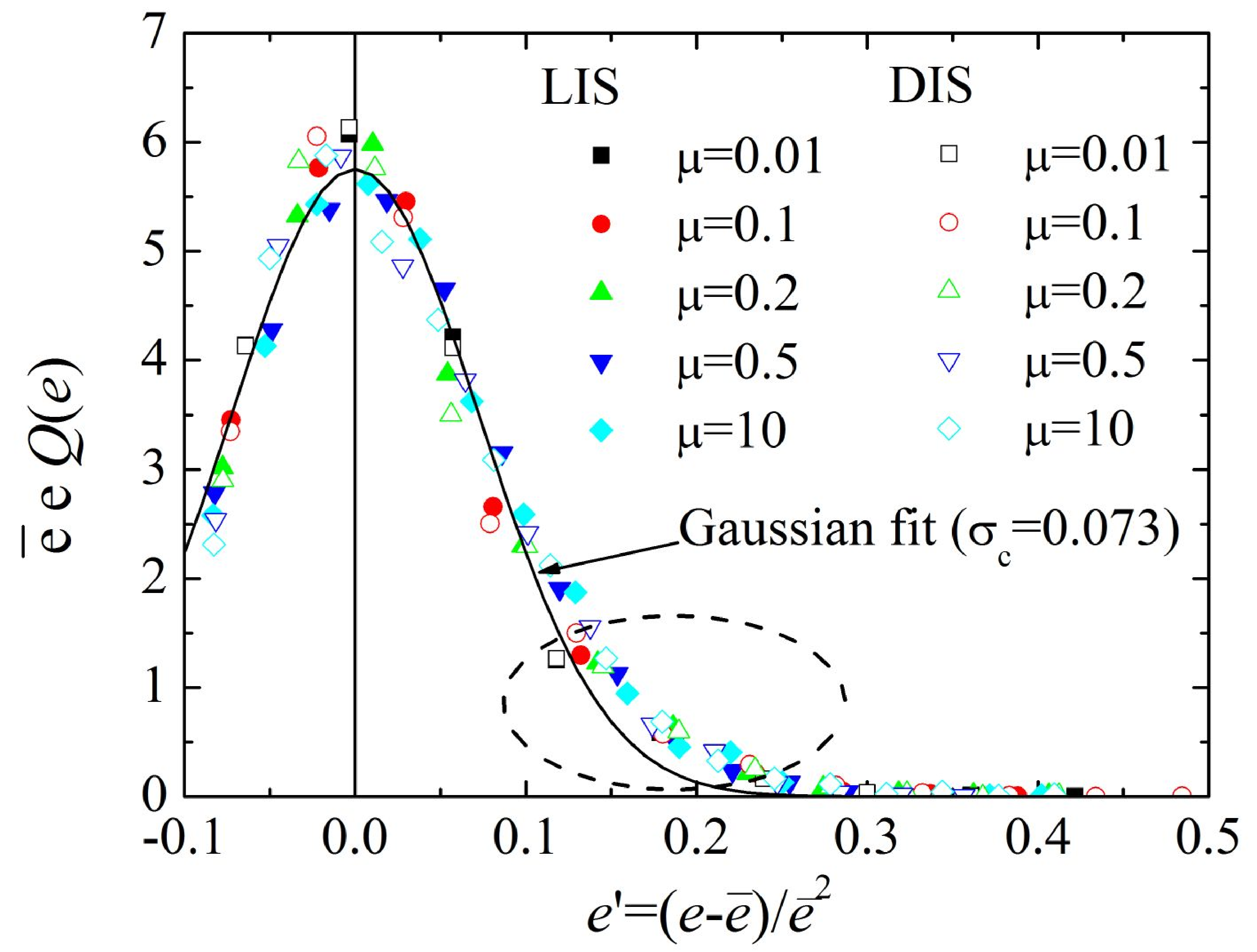}
\caption{The normalized cell probability $\bar{e}e Q(e)$. It is fitted well by a Gaussian form, except for small, but consistent, deviations in the large-$e$ tail (highlighted by a dash-lined ellipse).}
\label{fig:normalized_qe}
\end{minipage}
\end{figure} 

\subsection{\label{sec4-3} Conditional quadron volume distributions}

\begin{figure}[!bht]
\begin{minipage}[t]{0.4\textwidth}
\includegraphics[width=1.0\textwidth]{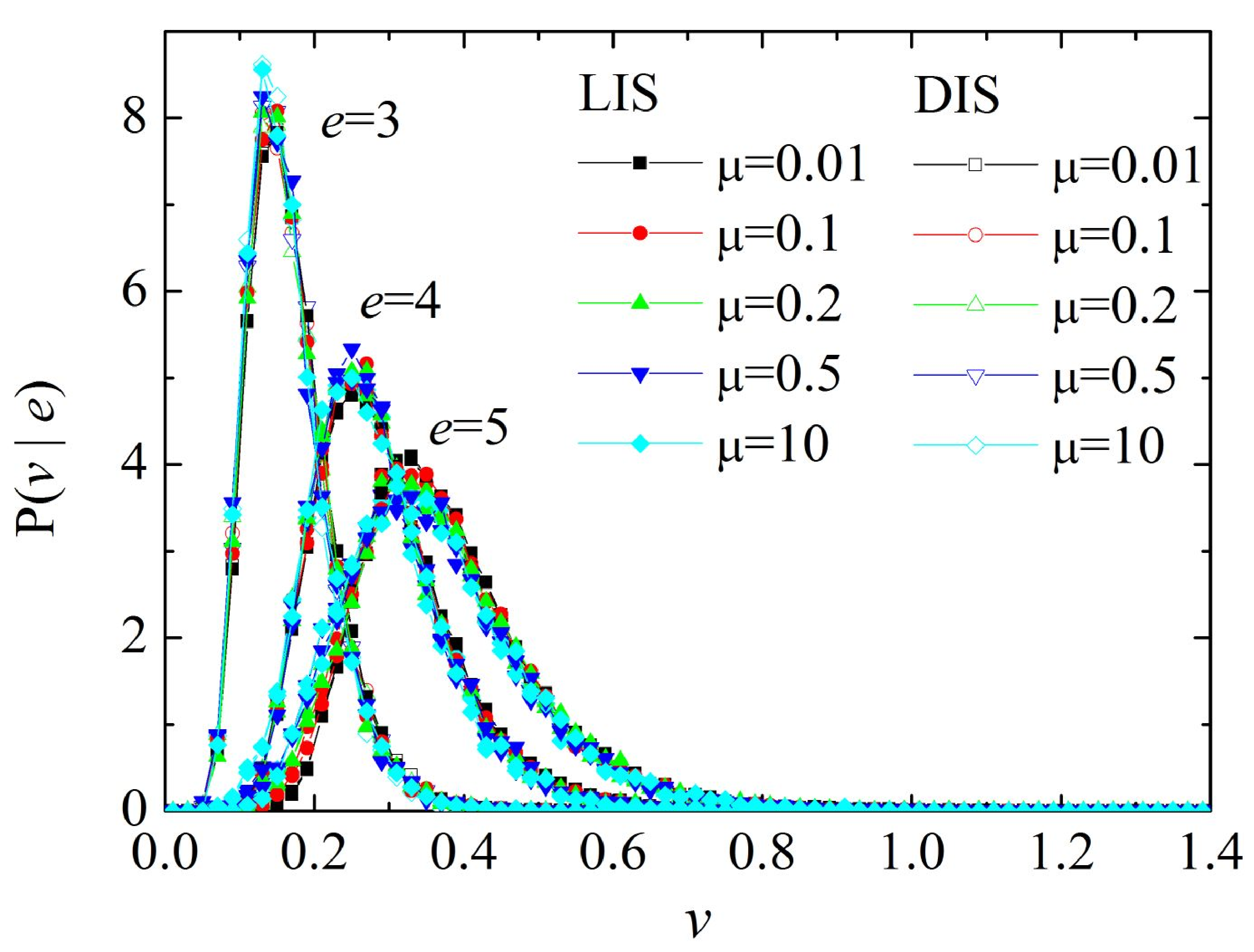}
\caption{The conditional PDF of the quadron volumes for $e \le 5$ collapse nicely for all the $\mu$ and initial conditions}.
\label{fig:cPDF_e5}
\end{minipage}
\end{figure}

\begin{figure}[!bht]
\begin{minipage}[t]{0.4\textwidth}
\includegraphics[width=1.0\textwidth]{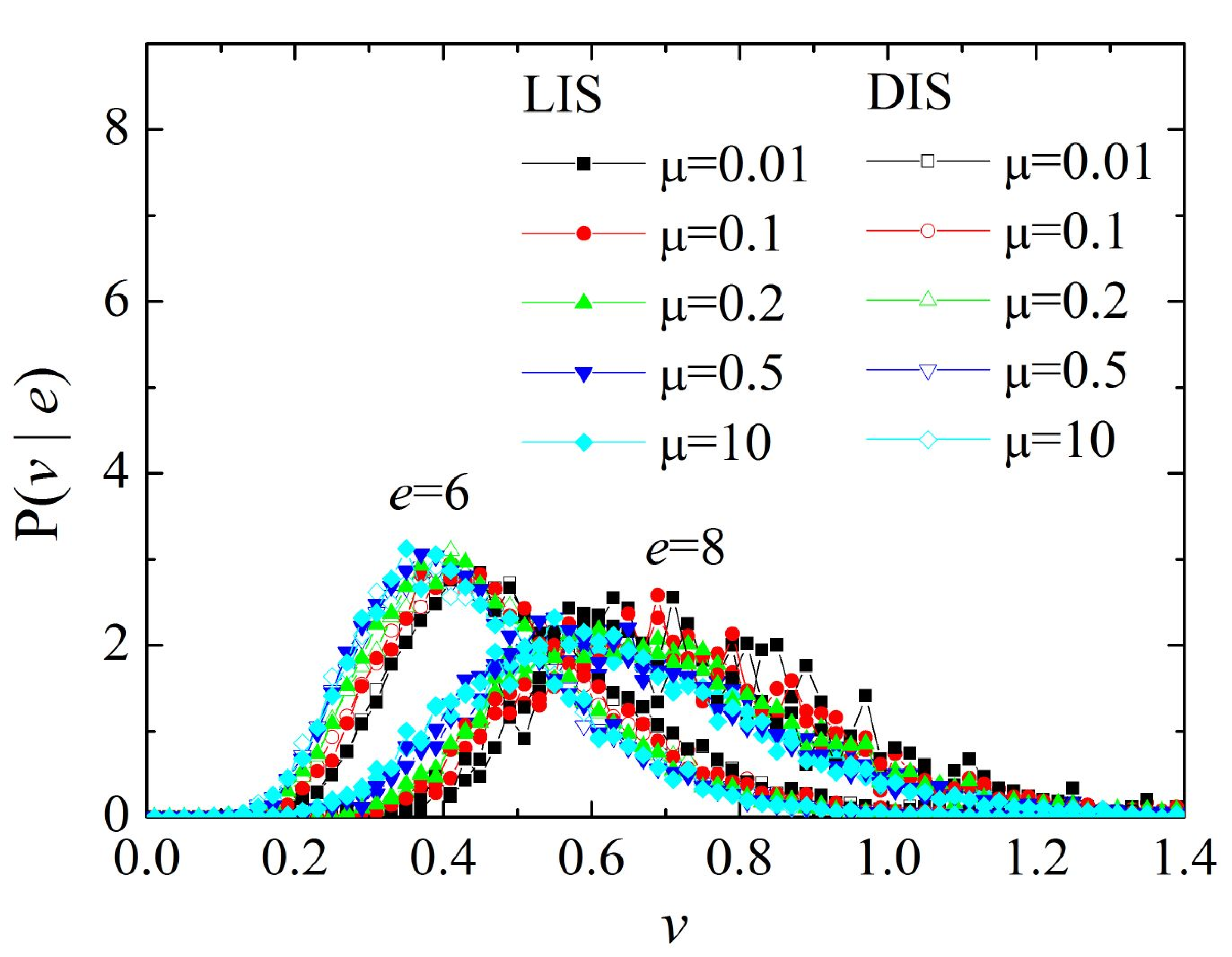}
\caption{The conditional PDF of the quadron volumes for $e = 6, 8$ collapse, but not as sharply as those for $e \le 5$ (Fig. \ref{fig:cPDF_e5}). This is a result of the underrepresentation of mechanically unstable elongated large cells, which suppresses occurrence of small volume quadrons.}
\label{fig:cPDF_e6}
\end{minipage}
\end{figure}

\begin{figure}[!bht]
\begin{minipage}[t]{0.4\textwidth}
\includegraphics[width=1.0\textwidth]{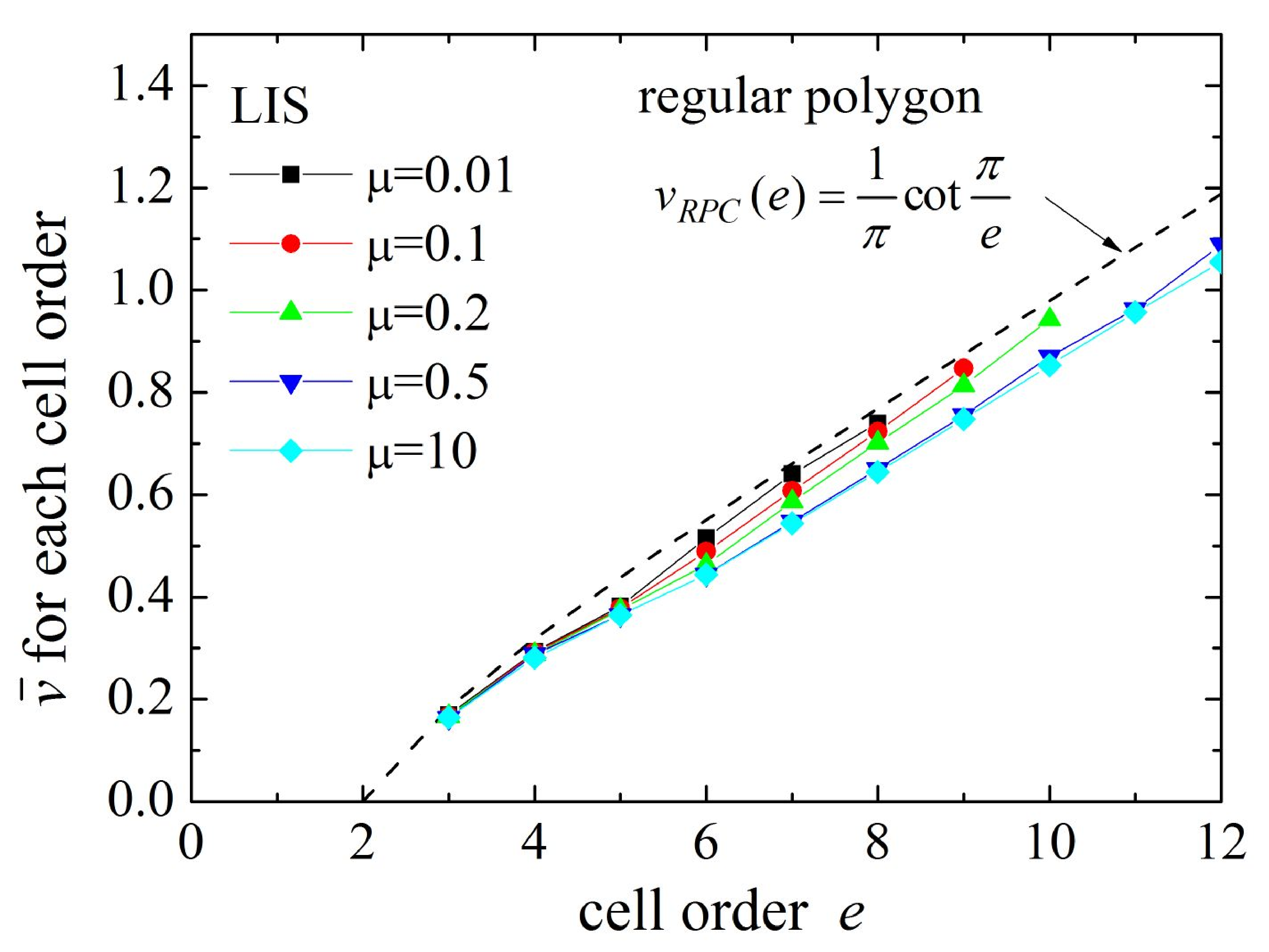}
\caption{The variation of the mean conditional quadron volume, $\bar{v}$, with cell order $e$. The quadron volume of the regular polygon cell  (dashed line) form an upper bound.}
\label{fig:Mean Vq by cell order}
\end{minipage}
\end{figure}

\begin{figure}[!bht]
\begin{minipage}[t]{0.4\textwidth}
\includegraphics[width=1.0\textwidth]{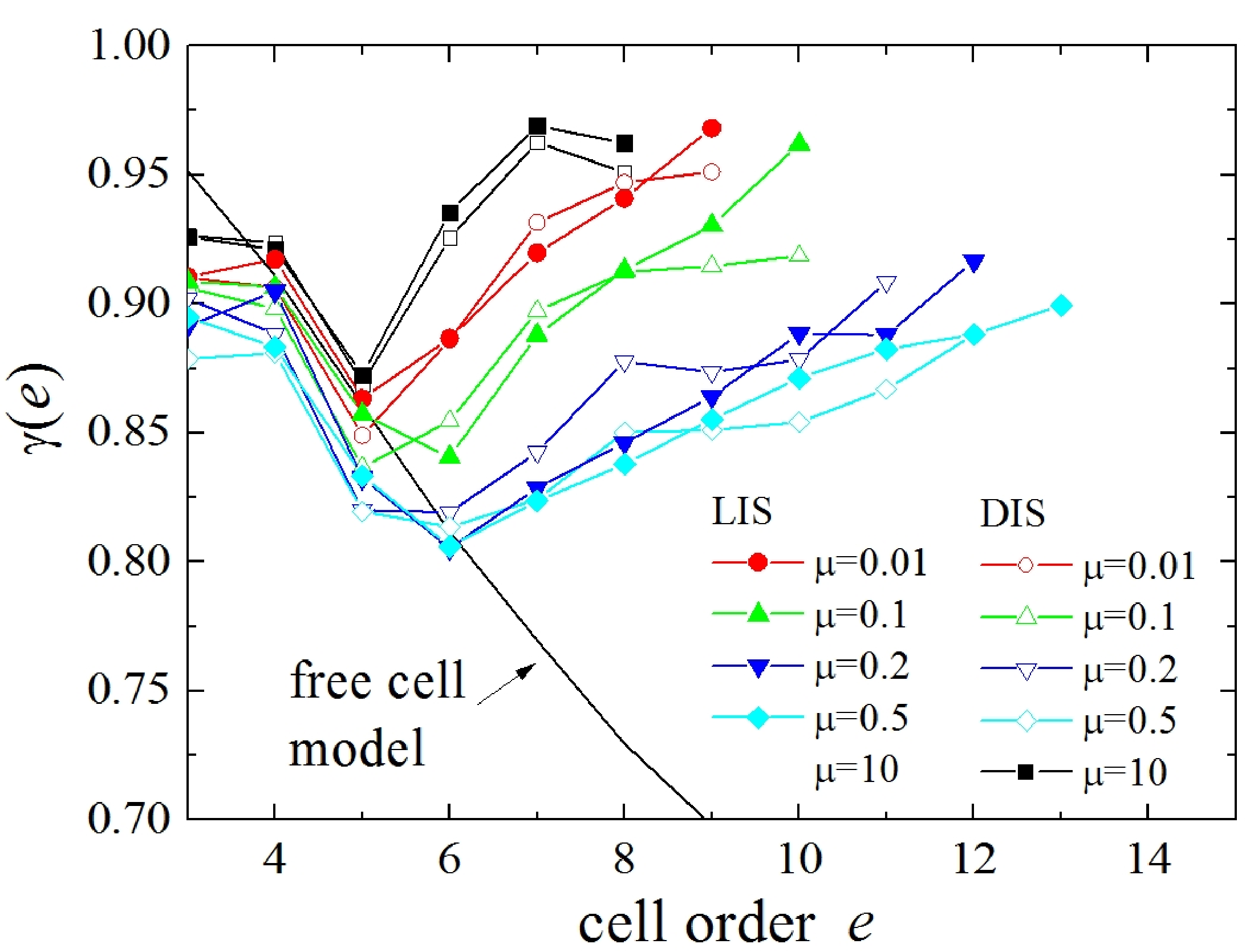}
\caption{The ratio of the mean quadron volume to the quadron volume of regular polygon, $\gamma$ ($<1$), plotted as a function of cell order $e$.
Note that the FCM describes well the geometric effect leading to the decrease in $\gamma$ for $e<5$, but cannot show the increase at larger values since it does not include effects of mechanical stability.}
\label{fig:vq_vrp}
\end{minipage}
\end{figure}

To gain further insight into the universal properties of the structure, let us consider in more detail the conditional PDFs $P\left(v\mid e\right)$. 
These PDFs were studied by Frenkel et al. \cite{Fretal08}, who argued, on the basis of geometrical considerations, that they should be independent of inter-granular friction. Their argument was based on the observation that, given a collection of $N$ arbitrary grains, the number of ways to arrange $e$ grains into a cell of order $e$  depends only on the grains shapes and not on the inter-granular  friction. 
While our results seem to provide a support to this argument, a closer look shows a systematic $\mu$-dependence of $P\left(v\mid e\right)$ for $e> 6$ (Fig. \ref{fig:cPDF_e5} and \ref{fig:cPDF_e6}). We can also see this effect in the behaviour of the mean of the conditional quadron volume as a function of $e$, $\bar{v}(e)=\sum_{q\in e}v P\left(v\mid e\right)$, shown in Fig. \ref{fig:Mean Vq by cell order}. We include in the figure calculated values of quadron volumes of regular polygonal cells (RPC) of order $e$, 
\begin{eqnarray}
v^{RPC}(e) = \frac{V_q^{RPC}(e)}{\bar{V}_g} = \frac{1}{\pi} cot \frac{\pi}{e}
\label{eq:regular_polygon}
\end{eqnarray}
which is clearly an upper bound for $\bar{v}(e)$. 
It is constructive to consider the ratio of the observed mean quadron volume to the regular polygon value, $\gamma(e)\equiv\bar{V}_q(e)/\bar{V}_q^{RPC}(e)$, shown in Fig. \ref{fig:vq_vrp}. We observe that $\gamma(e)$, which is always below unity, has a minimum between $e=5$ and $e=6$ and that it decreases with $\mu$. These can be understood as follows. Carrying out a cell shape analysis (to be reported elsewhere), where cells were approximated to lowest order as ellipses, there is a strong correlation between $\gamma(e)$ and the mean aspect ratio of the ellipses for any given $e$.  
Our interpretation is that the increase of $\gamma(e)$ with $\mu$ is due to the decrease of mechanical stability of elongated cells as $\mu$ decreases.
The behaviour for any particular $\mu$ depends then on two effects. As $e$ increases, more cell configurations can be realised, allowing for more elongated cells and this accounts for the initial decrease of $\gamma(e)$ with $e$.
However, the more elongated the cell, the less stable it is, leading to fewer elongated cells as $e$ increases. 
The competition between the increase in number of possible configurations and limiting stability constraints gives rise to the observed minimum.
To test this understanding, we use next an analytical, free cell model (FCM).


\section{The free cell model}

\begin{figure}[!bht]
\begin{minipage}[t]{0.4\textwidth}
\includegraphics[width=1.0\textwidth]{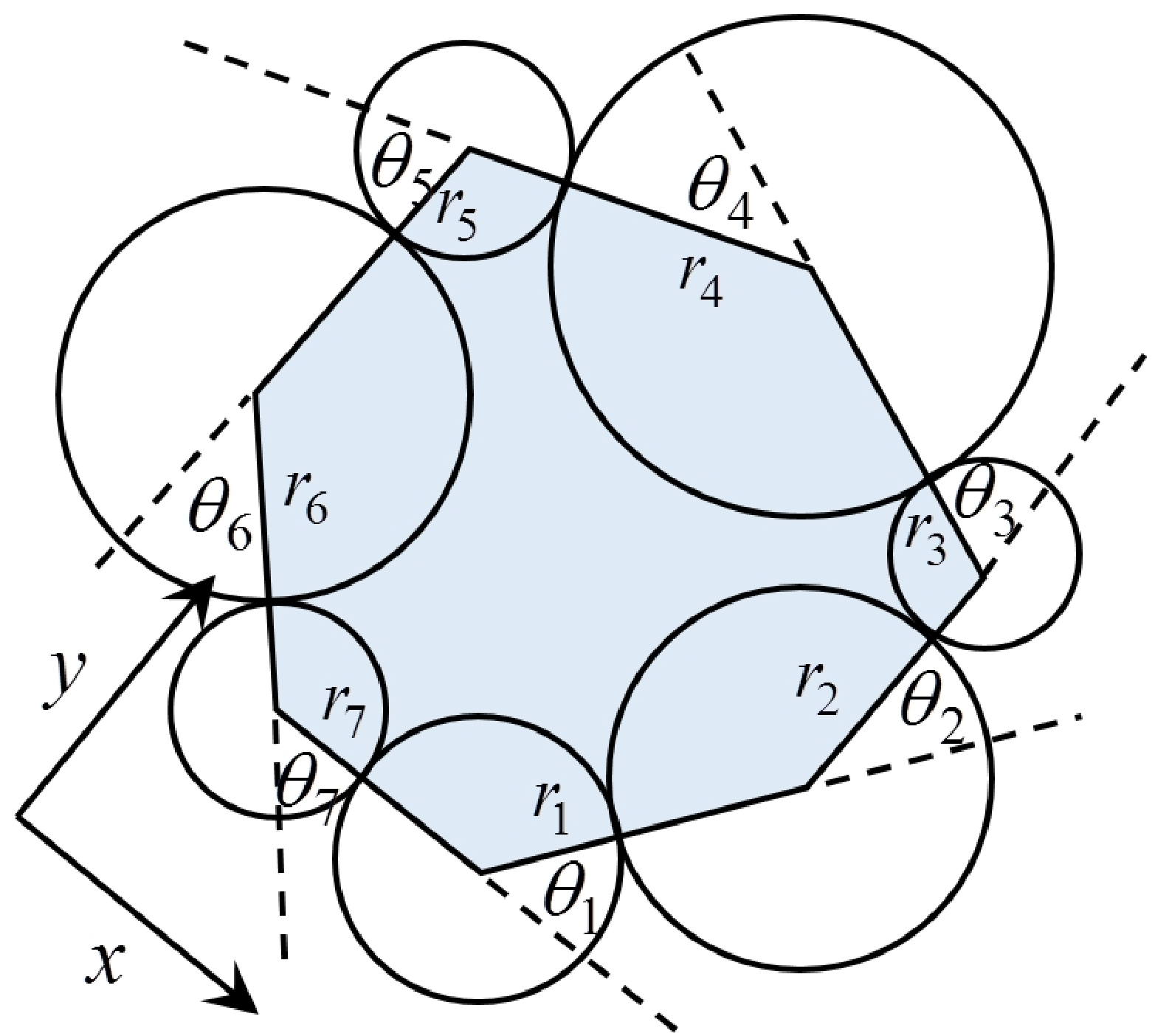}
\caption{The construction of a sample cell configuration generated by the free cell model.}
\label{fig19}
\end{minipage}
\end{figure}

\begin{figure}[!bht]
\begin{minipage}[t]{0.4\textwidth}
\includegraphics[width=0.8\textwidth]{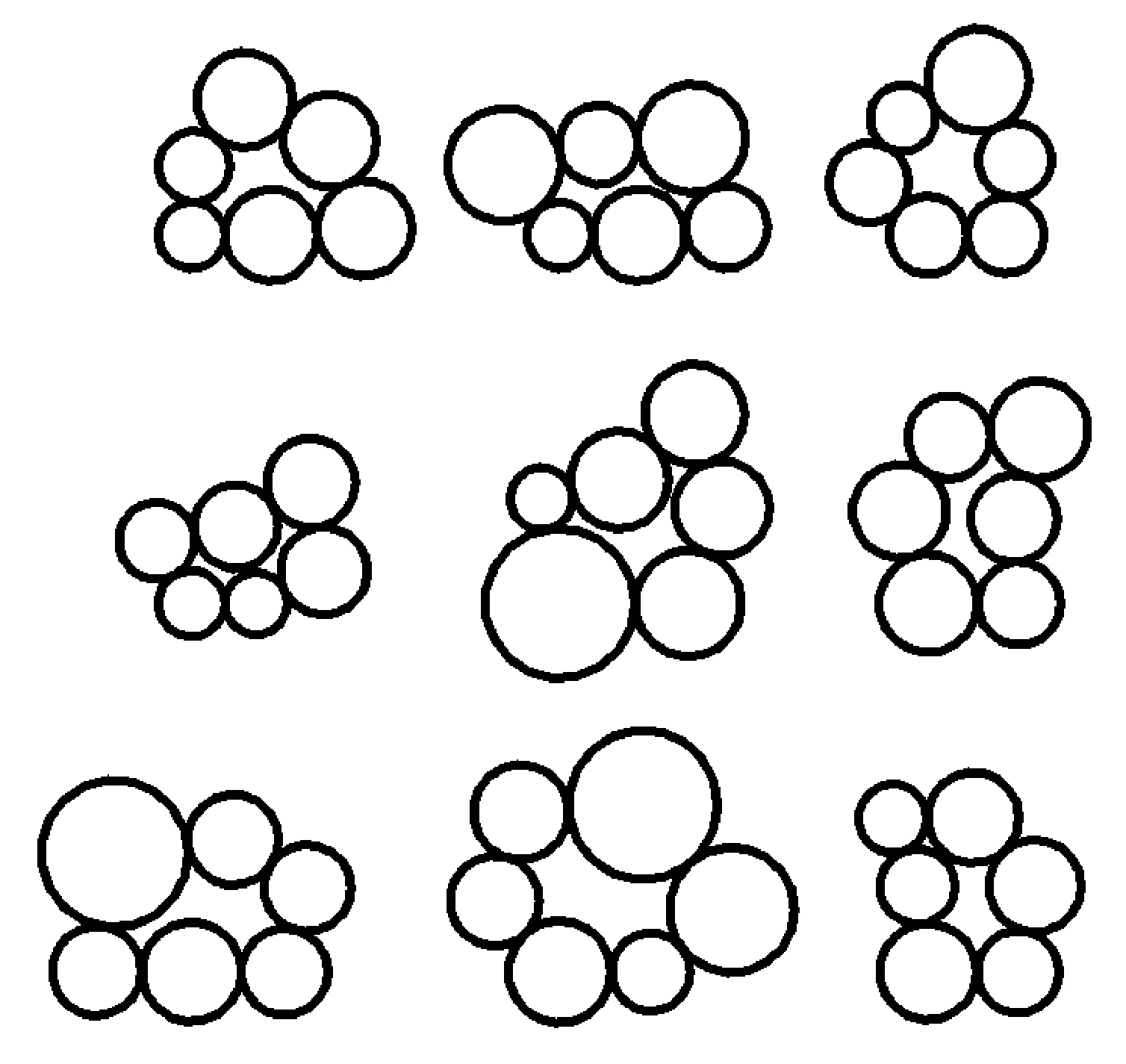}
\caption{ Examples of free cell configurations for $e=6$.}
\label{fig20}
\end{minipage}
\end{figure}

The free cell is a single closed cell of order $e$, whose surrounding discs are chosen randomly from the same size distribution as in the above DEM simulations. 
The model consists of generating a large number of such cells and analysing the resulting quadron volumes distribution. 
The cell is constructed as follows. First, a random set of $e$ discs is generated from the given size distribution. Then a random set of $e$ angles is generated 
($\theta_1$ to $\theta_6$ in Fig. \ref{fig19}), satisfying

\beq
\sum_{j=1}^{e} \theta_j = 2\pi 
\label{eq:fcm01}
\eeq 
and
\beq
\sum_{j=1}^{e} {x_j\choose y_j} = \sum_{j=1}^{e} (r_j+r_{j+1}) {\cos{\omega_j}\choose\sin{\omega_j}} = 0
\label{eq:fcm02}
\eeq 
where $x_j$ and $y_j$ are the coordinates of the centre of the $j$th disc, $r_{e+1}=r_1$ and $\omega_j\equiv \sum_{k=1}^j \theta_k$.
When $e=3$, the three angles, $\theta_j$ can be determined uniquely for given three radii $r_j$, but when $e > 3$, the angles are under-determinated. We then fix $(e-3)$ angles randomly and obtain the rest by solving the above equations, using the Newton-Raphson iteration scheme. We check that the solution satisfies the condition that disc $j$ has contacts with discs $(j-1)$ and $(j+1)$. Finally, we check that each disc has no additional contact or overlap with any of the other discs, otherwise the solution is discarded and a new cell is generated.
Fig. \ref{fig20} shows examples of free cells generated by this procedure. 
The FCM allows us to generate quickly and efficiently a large ensemble of cells, whose statistics can be compared with the DEM results for the purpose of analysing the effects of mechanical stability on the cell shape distribution. 

\begin{figure}[!bht]
\begin{minipage}[t]{0.4\textwidth}
\includegraphics[width=1.0\textwidth]{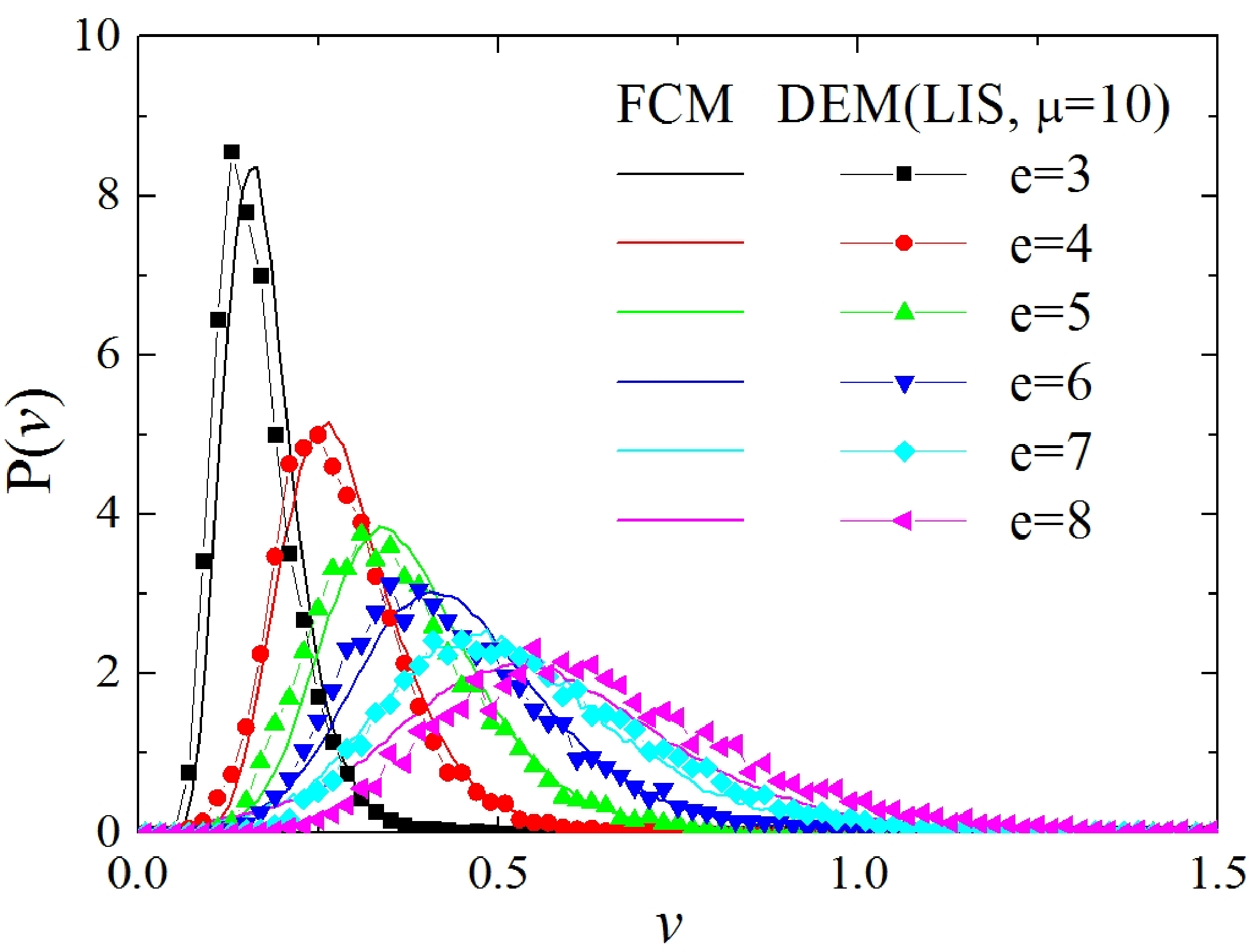}
\caption{A comparison of the PDF of quadron volumes, generated both with the free cells model and with the DEM numerical experiments.}
\label{fig21}
\end{minipage}
\end{figure}

In Fig. \ref{fig21} we compare the resulting conditional quadron volume PDFs from the FCM and DEM simulations for $\mu=10$ and $e$=3 -- 8. 
For $e\le 6$ the PDFs are very close, but the mean quadron volume is consistently lower in the FCM, with the difference increasing as $e$ increases.
That the two PDFs do not coincide perfectly for any value of $e$ is a result of three effects. 
One is our observation that the size distribution of discs surrounding cells of order $e$ in the DEM simulations depends, albeit very weakly, on $e$. This interesting effect is somewhat tangential to the thrust of this paper and is left to a later investigation. 
Another effect is that, in the simulations, as in real systems, the centroid of a disc contacts need not coincide with its geometric centre, which constitutes one of the quadron vertices in the FCM. This is because the FCM discs have only two contacts each and the disc centre is the only reasonable choice to make.  

A careful analysis of these two effects shows that their combined contribution is negligible compared with the effect of mechanical stability constraint. The condition that the cell be stable under compressive forces was not taken into consideration by Frenkel et al. \cite{Fretal08}. 
Our DEM simulations show that this constraint biases the distribution towards a larger proportion of cells with aspect ratios close to one compared with the FCM. 
Indeed, the higher the intergranular friction coefficient, the more elongated can cells be and still be stable. The effect of mechanical instability gets stronger with decreasing friction coefficient and large cell orders. Since elongated cells tend to have, on average, smaller quadron volumes, the instability suppresses the small volume tail of the distribution.
This effect is captured well by the FCM. 
Fig. \ref{fig:vq_vrp} includes the ratio, $\gamma$, of the mean quadron volume, generated by the FCM, to the quadron volume of regular polygon. It mimics well the effect of the the decrease in $\gamma$ for $e<5$ and for $e<6$, for low and high friction coefficients, respectively. It also demonstrates that the increase in $\gamma$ for higher values of $e$ in the DEM is a direct result of the mechanical stability constraint, which is not included in, and therefore not captured by, the FCM.

\section{The collapse and the uniform measure assumption}

A significant assumption, underlying much of the work on GSM, is that of uniform measure, i.e. that any possible microstate, structural and stress configuration, can be realised in the ensemble. In other words, that the occurrence probability of a microstate depends only on the Boltzmann-like factor in the partition function, in our case the exponential term in eq. (\ref{Z1}), and that it depends neither on the generating dynamics of the systems of the ensemble nor on their sampling procedure. In general, the partition function must also include a `measure', i.e. a probability density function, $G$, that specifies the probability such dynamics and/or sampling. Generally, this function may depend on both the degrees of freedom and on the details of the systems generation dynamics and sampling process. The term uniform measure refers to $G$ being constant. It was because of the difficulty to parameterise the dynamics and sampling, that the uniform measure was assumed originally for granular matter \cite{EdOa89b} for simplicity.

This assumption became a subject of some scrutiny in recent years \cite{xu2005random,gao2006frequency,gao2007enumeration,gao2009geometrical,gao2009experimental,Heetal07,PaFr12, Bletal15}. It has been shown to hold in some systems \cite{Heetal07} and to fail in others \cite{Fretal13,Pa15}. 
Focusing on this assumption for the structural sub-ensemble, the collapses we observed of several distributions of structural characteristics, although not resolving this issue generally, provide good insight into the processes that would give rise to a non-uniform measure. Specifically, these collapses reveal an underlying type of order that can only be the result of the self-organisation of the granular medium during the dynamic process we used to generate them. The existence of such regularities means that the structures of all the systems we studied are not entirely random, namely, there is a large class of completely random systems that are not represented in the ensemble and which are side-lined by the generation processes. Clearly, some of these configurations have a Boltzmann-like factor of exactly the same value as the more frequent configurations we observed. It follows that not all configurations with the same Boltzmann factor are equally probable and, therefore, that the measure cannot be uniform when the generation process leads to self-organised structures. 
Now, it seems plausible that almost all mechanically equilibrated granular packings, assembled under external forces, would self-organise in some fashion. This argument implies that the assumption of uniform measure must fail in all, or almost all such ensembles. 
This also means that, to derive realistic equations of state and constitutive relations from GSM, which is the ultimate goal of any statistical mechanics, one must have good models for the measure function $G$.
\\

\section{Conclusions and Discussions}

To conclude, we have studied in detail the effects of inter-granular friction and initial conditions on the structural characteristics of 2D granular assemblies. We have shown that structure can be analysed quantitatively, using the quadron description\cite{BaBl02,BlEd03}. In particular, we have established that a number of distributions collapse onto single curves, pointing to a `universal'-like structural characteristics that are independent not only of initial state, but also of the friction coefficient. These are significant results, whose implications are discussed at the end of this section.

The raw quadron volume distribution, which is key for the Edwards formulation of volumetric granular statistical mechanics \cite{EdOa89a,EdOa89b,BlEd03,BlEd06,HiBl12}, changes systematically towards larger values with increasing friction coefficient $\mu$ and hence with decreasing mean coordination number $\bar{z}$. However, when normalised by the mean quadron volume, it collapses nicely onto a single curve. Following the insight proposed by Frenkel et al.\cite{Fretal08}, we have traced this to a similar collapse of the conditional PDFs $P(V_q/\bar{V}_g\mid e)$ for every cell order $e$, except for small, but insightful, deviations for large $e$, which are discussed below.

Suggestions that there is a relation between the mean coordination number, $\bar{z}$ and the packing fraction date back to 1929 \cite{smith1929packing}, but later studies showed that such a relation depends also on the initial conditions \cite{Mitchell-Soga-2005,silbert2010,katagiri2014variations}. The idea that an initial-state-independent relation could exist if rattlers are disregarded, was implied, but not stated as such, in \cite{Meetal10}. The collapse we observed \cite{MaBl14} of the quadron volume conditional distributions, which disregards rattlers, supports this view. However, we found in \cite{MaBl14} that the model proposed in \cite{Meetal10} leads to an incorrect $\bar{z}-\phi'$ relation and our analysis, described in more detail here, agrees better with the numerical simulations of frictional particles presented here.
This is consistent with our observation that the plot of $\bar{V}_q$ as a function of $\bar{z}$ is also independent of the initial state.
Since rattlers do not support forces in our systems, these results carry a fundamental significance - they suggest that the packing fraction, the mean coordination number, the quadron volume distribution and generally the structure are linked directly to the manner in which stresses are transmitted.
While this gives insight into the structure of the system, it should be remembered that it is difficult to estimate the packing fraction of rattlers in real experiments and it is unclear at this stage how to use this insight for engineering applications.

Considerations of mechanical stability also suggest that the argument proposed by Frenkel at el.\cite{Fretal08}, leading to independence of the conditional quadron volume distributions of inter-granular friction, needs to be modified. Their argument assumes that cell shapes are independent of mechanics and therefore that arranging an $e$-sided cell depends only on the grain shape distribution. This, in turn, eliminates friction as a factor in the occurence probability of cell structures. However, as friction decreases,  elongated cells are less stable mechanically, reducing the number of available cell shapes. Indeed, we can observe a small increase in the occurrence probability of large quadron volumes in $P(V_q/\bar{V}_g\mid e)$ for $e>6$ as $\mu$ increases (Fig. \ref{fig:cPDF_e6}). 

This effect is seen most clearly when plotting the ratio of the mean quadron volume to that of a regular polygon as a function of $e$ (Fig. \ref{fig:vq_vrp}). The initial drop reflects the departure of the cell from a regular polygon due to the increased number of possible geometrical configurations, while the subsequent increase shows the effect of mechanical stability in limiting such a departure. The more pronounced increase for lower $\mu$ is evidence for our argument above.

\begin{figure}[!bht]
\begin{minipage}[t]{0.4\textwidth}
\includegraphics[width=1.0\textwidth]{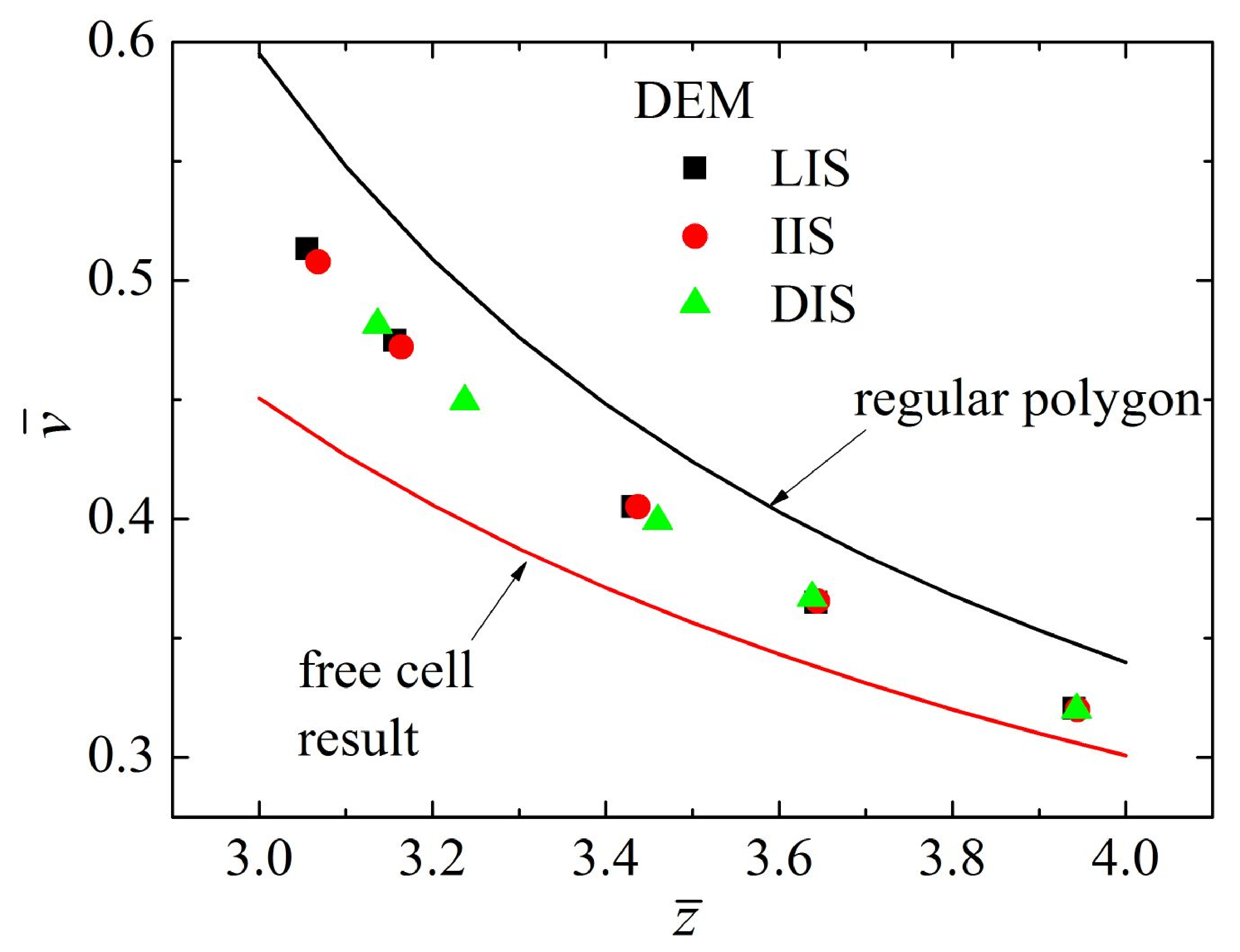}
\caption{Simple prediction of the mean quadron volume in terms of the mean coordination number.}
\label{fig:VqvsZg}
\end{minipage}
\end{figure}

To  illustrate the effect of mechanical stability, we have constructed a free cell model (FCM) that takes into account only geometric considerations and disregards mechanical stability. Indeed, this model shows the same initial decrease of the above ratio with $e$ and none of the subsequent increase (Fig. \ref{fig:vq_vrp}). 
Thus, neglecting the mechanical stability effect, the FCM provides a low bound for the mean quadron volume, as illustrated in Fig. \ref{fig:VqvsZg}.

We have further found that the raw cell order PDF, $Q(e)$, changes systematically with $\mu$, but hardly at all with the initial state. 
However, all the PDFs also collapse to a single curve under the  transformations $Q(e)\to\bar{e}eQ(e)$ and $e\to\left(e-\bar{e}\right)/\bar{e}^2$.  
We note that the collapse cannot be perfect at the large $e$ tail due to the aforementioned constraint of mechanical stability. This friction-specific behaviour, which is evidenced by Figs. \ref{fig:cPDF_e6} and \ref{fig:vq_vrp}, is hardly noticed at all in Fig. \ref{fig:normalized_qe}, demonstrating that the effect of mechanical stability is small compared to the geometrical effect discussed by Frenkel et al. \cite{Fretal08}.

Our results are significant for several reasons. 
First, they are a step towards a systematic understanding of structural organisation of granular matter in response to a specific packing procedure. In particular, they suggest that the structural characteristics of granular matter are best understood via the statistics of cell configurations. 
Second, they make it possible to progress on the statistical mechanics of granular matter, where knowledge of the distribution of the inter-contact vectors, $\vec{r}^q$, is key. 
Third, the results shed light on the common wisdom in the soil mechanics and civil engineering communities that the initial state and inter-granular friction affect the final relation between $\bar{z}$ and $\phi$ (\cite{Mitchell-Soga-2005,silbert2010,katagiri2014variations}). 
Our findings show that this relation is directly linked to mechanical stability of the packing backbone - for any specific pack generation process, the initial state and friction affect the fraction of rattlers, which do not participate in the stress transmission. Once the rattlers are disregarded, there emerges a unique relation between $\bar{z}$ and the corrected rattlers-free packing fraction $\phi'$.

Fourth, our analysis has a significant ramification for the famous packing problem. Different packing procedures lead to different structures. For example, procedures not constrained by mechanical stability would give rise to different $Q(e)$'s than those that are. Specifically, the former would tend to have lower packing fractions due to a higher occurrence probability of large cells, which are mechanically unstable. This reinforces the need for a more accurate definition of the packing problem, which we discuss elsewhere\cite{BlToMaSoon}.

Taken together, these results show that the packing process has a fundamental effect on the structural characteristics while the initial conditions and the inter-granular friction are details that only modify this effect and can be scaled away. The collapsed distributions show how to predict such modifications.
It should be emphasised that the collapses, which we have found, do not imply universality in the traditional sense because the specific forms of the collapsed curves depend on the packing procedure. 
For example, it has been demonstrated that, using a different packing protocol, it is possible to generate packs dominated by cells of order 3 \cite{HiBl12}, which cannot be scaled to match the ones we obtained here.  There is in principle an infinite number of possible procedures to generate packs, depending on an astronomically large number of parameters. It would be impossible to map all the possible processes to a manageable parameter space. Each such procedure gives rise to its own characteristic $Q(e)$ and it would be impossible to collapse all of these $Q(e)$ onto one curve. 

Our results have an intriguing and potentially significant implication for the much discussed issue of uniform measure in granular statistical mechanics\cite{xu2005random,gao2006frequency,gao2007enumeration,gao2009geometrical,gao2009experimental,Heetal07,PaFr12,Fretal13, Bletal15, Pa15}. We argued that the collapse of distributions of various structural characteristics points to self-organisation under external loading and, therefore, to under-representation of many random structural configurations in the ensemble. This means that, at least in our systems, and possibly in all ensembles of systems brought to mechanical equilibrium under external forces, the assumption of uniform measure should fail. We believe that most, if not all, such systems self-organise into their final structure and, therefore, that our results can be interpreted to suggest a widespread failure of the uniform measure assumption. If this conjecture is supported in future studies then work must focus on deriving models of the measure function $G$, if we are to make progress on applying GSM to derivation of equations of state.

We argued that the structural features studied here should be very little affected, if at all, by the stress microstates and the angoricity \cite{BlEd05,Heetal07,BlEd09} and our results appear to support this argument. Unavoidably, the different systems and samples we studied could not be generated under exactly the same distribution of external forces and each configuration of stress microstate could, in principle, affect the partition function (\ref{Z1}). Yet, the collapses we observed appeared to be sample-independent, suggesting insensitivity of the self-organised structural characteristics to the angoricity.

To fully understand how structural characteristics depend on the packing procedure, one needs a good model of both the geometric effects and the limitations that mechanical stability poses on the cell shape distribution. We are currently developing such a model and will report results in a later publication.
Intriguingly, the relevance of both structural and stress effects is reminiscent of the inter-dependence demonstrated for the statistical mechanical understanding of granular matter \cite{Bletal12} and it would be interesting to find out whether the two issues are related or not.

\nocite{*}

 \bibliography{cell_geometry}

\begin{thebibliography}{47}%
\makeatletter
\providecommand \@ifxundefined [1]{%
 \@ifx{#1\undefined}
}%
\providecommand \@ifnum [1]{%
 \ifnum #1\expandafter \@firstoftwo
 \else \expandafter \@secondoftwo
 \fi
}%
\providecommand \@ifx [1]{%
 \ifx #1\expandafter \@firstoftwo
 \else \expandafter \@secondoftwo
 \fi
}%
\providecommand \natexlab [1]{#1}%
\providecommand \enquote  [1]{``#1''}%
\providecommand \bibnamefont  [1]{#1}%
\providecommand \bibfnamefont [1]{#1}%
\providecommand \citenamefont [1]{#1}%
\providecommand \href@noop [0]{\@secondoftwo}%
\providecommand \href [0]{\begingroup \@sanitize@url \@href}%
\providecommand \@href[1]{\@@startlink{#1}\@@href}%
\providecommand \@@href[1]{\endgroup#1\@@endlink}%
\providecommand \@sanitize@url [0]{\catcode `\\12\catcode `\$12\catcode
  `\&12\catcode `\#12\catcode `\^12\catcode `\_12\catcode `\%12\relax}%
\providecommand \@@startlink[1]{}%
\providecommand \@@endlink[0]{}%
\providecommand \url  [0]{\begingroup\@sanitize@url \@url }%
\providecommand \@url [1]{\endgroup\@href {#1}{\urlprefix }}%
\providecommand \urlprefix  [0]{URL }%
\providecommand \Eprint [0]{\href }%
\providecommand \doibase [0]{http://dx.doi.org/}%
\providecommand \selectlanguage [0]{\@gobble}%
\providecommand \bibinfo  [0]{\@secondoftwo}%
\providecommand \bibfield  [0]{\@secondoftwo}%
\providecommand \translation [1]{[#1]}%
\providecommand \BibitemOpen [0]{}%
\providecommand \bibitemStop [0]{}%
\providecommand \bibitemNoStop [0]{.\EOS\space}%
\providecommand \EOS [0]{\spacefactor3000\relax}%
\providecommand \BibitemShut  [1]{\csname bibitem#1\endcsname}%
\let\auto@bib@innerbib\@empty
\bibitem [{\citenamefont {Matsushima}\ and\ \citenamefont
  {Blumenfeld}(2014)}]{MaBl14}%
  \BibitemOpen
  \bibfield  {author} {\bibinfo {author} {\bibfnamefont {T.}~\bibnamefont
  {Matsushima}}\ and\ \bibinfo {author} {\bibfnamefont {R.}~\bibnamefont
  {Blumenfeld}},\ }\href@noop {} {\bibfield  {journal} {\bibinfo  {journal}
  {Phys. Rev. Lett.}\ }\textbf {\bibinfo {volume} {112}},\ \bibinfo {pages}
  {098003} (\bibinfo {year} {2014})}\BibitemShut {NoStop}%
\bibitem [{\citenamefont {Bernal}(1960)}]{Bernal1960}%
  \BibitemOpen
  \bibfield  {author} {\bibinfo {author} {\bibfnamefont {J.}~\bibnamefont
  {Bernal}},\ }\href@noop {} {\bibfield  {journal} {\bibinfo  {journal}
  {Nature}\ }\textbf {\bibinfo {volume} {185}},\ \bibinfo {pages} {68}
  (\bibinfo {year} {1960})}\BibitemShut {NoStop}%
\bibitem [{\citenamefont {Mogami}(1965)}]{Mogami1965}%
  \BibitemOpen
  \bibfield  {author} {\bibinfo {author} {\bibfnamefont {T.}~\bibnamefont
  {Mogami}},\ }\href@noop {} {\bibfield  {journal} {\bibinfo  {journal} {Soils
  and Foundations}\ }\textbf {\bibinfo {volume} {5(2)}},\ \bibinfo {pages} {26}
  (\bibinfo {year} {1965})}\BibitemShut {NoStop}%
\bibitem [{\citenamefont {Ball}\ and\ \citenamefont
  {Blumenfeld}(2002)}]{BaBl02}%
  \BibitemOpen
  \bibfield  {author} {\bibinfo {author} {\bibfnamefont {R.~C.}\ \bibnamefont
  {Ball}}\ and\ \bibinfo {author} {\bibfnamefont {R.}~\bibnamefont
  {Blumenfeld}},\ }\href@noop {} {\bibfield  {journal} {\bibinfo  {journal}
  {Phys. Rev. Lett.}\ }\textbf {\bibinfo {volume} {88}},\ \bibinfo {pages}
  {115505} (\bibinfo {year} {2002})}\BibitemShut {NoStop}%
\bibitem [{\citenamefont {Blumenfeld}\ and\ \citenamefont
  {Edwards}(2003)}]{BlEd03}%
  \BibitemOpen
  \bibfield  {author} {\bibinfo {author} {\bibfnamefont {R.}~\bibnamefont
  {Blumenfeld}}\ and\ \bibinfo {author} {\bibfnamefont {S.}~\bibnamefont
  {Edwards}},\ }\href@noop {} {\bibfield  {journal} {\bibinfo  {journal} {Phys.
  Rev. Lett.}\ }\textbf {\bibinfo {volume} {90}},\ \bibinfo {pages} {114303}
  (\bibinfo {year} {2003})}\BibitemShut {NoStop}%
\bibitem [{\citenamefont {Blumenfeld}\ and\ \citenamefont
  {Edwards}(2006)}]{BlEd06}%
  \BibitemOpen
  \bibfield  {author} {\bibinfo {author} {\bibfnamefont {R.}~\bibnamefont
  {Blumenfeld}}\ and\ \bibinfo {author} {\bibfnamefont {S.}~\bibnamefont
  {Edwards}},\ }\href@noop {} {\bibfield  {journal} {\bibinfo  {journal} {Euro.
  Phys. J. E}\ }\textbf {\bibinfo {volume} {19}},\ \bibinfo {pages} {23}
  (\bibinfo {year} {2006})}\BibitemShut {NoStop}%
\bibitem [{\citenamefont {Blumenfeld}(2004)}]{Bl04}%
  \BibitemOpen
  \bibfield  {author} {\bibinfo {author} {\bibfnamefont {R.}~\bibnamefont
  {Blumenfeld}},\ }\href {\doibase 10.1103/PhysRevLett.93.108301} {\bibfield
  {journal} {\bibinfo  {journal} {Phys. Rev. Lett.}\ }\textbf {\bibinfo
  {volume} {93}},\ \bibinfo {pages} {108301} (\bibinfo {year}
  {2004})}\BibitemShut {NoStop}%
\bibitem [{\citenamefont {Gerritsen}\ \emph {et~al.}(2008)\citenamefont
  {Gerritsen}, \citenamefont {Kreiss},\ and\ \citenamefont
  {Blumenfeld}}]{Geetal08}%
  \BibitemOpen
  \bibfield  {author} {\bibinfo {author} {\bibfnamefont {M.}~\bibnamefont
  {Gerritsen}}, \bibinfo {author} {\bibfnamefont {G.}~\bibnamefont {Kreiss}}, \
  and\ \bibinfo {author} {\bibfnamefont {R.}~\bibnamefont {Blumenfeld}},\
  }\href@noop {} {\bibfield  {journal} {\bibinfo  {journal} {Physica A:
  Statistical Mechanics and its Applications}\ }\textbf {\bibinfo {volume}
  {387}},\ \bibinfo {pages} {6263} (\bibinfo {year} {2008})}\BibitemShut
  {NoStop}%
\bibitem [{\citenamefont {Blumenfeld}\ \emph {et~al.}(2012)\citenamefont
  {Blumenfeld}, \citenamefont {Jordan},\ and\ \citenamefont
  {Edwards}}]{Bletal12}%
  \BibitemOpen
  \bibfield  {author} {\bibinfo {author} {\bibfnamefont {R.}~\bibnamefont
  {Blumenfeld}}, \bibinfo {author} {\bibfnamefont {J.~F.}\ \bibnamefont
  {Jordan}}, \ and\ \bibinfo {author} {\bibfnamefont {S.~F.}\ \bibnamefont
  {Edwards}},\ }\href {\doibase 10.1103/PhysRevLett.109.238001} {\bibfield
  {journal} {\bibinfo  {journal} {Phys. Rev. Lett.}\ }\textbf {\bibinfo
  {volume} {109}},\ \bibinfo {pages} {238001} (\bibinfo {year}
  {2012})}\BibitemShut {NoStop}%
\bibitem [{\citenamefont {Blumenfeld}(2008)}]{Bl08a}%
  \BibitemOpen
  \bibfield  {author} {\bibinfo {author} {\bibfnamefont {R.}~\bibnamefont
  {Blumenfeld}},\ }in\ \href@noop {} {\emph {\bibinfo {booktitle} {Lecture
  Notes in Complex Systems}}},\ \bibinfo {editor} {edited by\ \bibinfo {editor}
  {\bibfnamefont {T.}~\bibnamefont {Aste}}, \bibinfo {editor} {\bibfnamefont
  {A.}~\bibnamefont {Tordesillas}}, \ and\ \bibinfo {editor} {\bibfnamefont
  {T.}~\bibnamefont {Matteo}}}\ (\bibinfo  {publisher} {World Scientific},\
  \bibinfo {address} {Singapore},\ \bibinfo {year} {2008})\ pp.\ \bibinfo
  {pages} {43--53}\BibitemShut {NoStop}%
\bibitem [{\citenamefont {Blumenfeld}\ and\ \citenamefont
  {Edwards}(2014)}]{BlEd14}%
  \BibitemOpen
  \bibfield  {author} {\bibinfo {author} {\bibfnamefont {R.}~\bibnamefont
  {Blumenfeld}}\ and\ \bibinfo {author} {\bibfnamefont {S.}~\bibnamefont
  {Edwards}},\ }\href {\doibase epjst/e2014-02258-y} {\bibfield  {journal}
  {\bibinfo  {journal} {Euro. Phys. J.}\ }\textbf {\bibinfo {volume} {223}},\
  \bibinfo {pages} {2189} (\bibinfo {year} {2014})}\BibitemShut {NoStop}%
\bibitem [{\citenamefont {Frenkel}\ \emph {et~al.}(2008)\citenamefont
  {Frenkel}, \citenamefont {Blumenfeld}, \citenamefont {Grof},\ and\
  \citenamefont {King}}]{Fretal08}%
  \BibitemOpen
  \bibfield  {author} {\bibinfo {author} {\bibfnamefont {G.}~\bibnamefont
  {Frenkel}}, \bibinfo {author} {\bibfnamefont {R.}~\bibnamefont {Blumenfeld}},
  \bibinfo {author} {\bibfnamefont {Z.}~\bibnamefont {Grof}}, \ and\ \bibinfo
  {author} {\bibfnamefont {P.}~\bibnamefont {King}},\ }\href@noop {} {\bibfield
   {journal} {\bibinfo  {journal} {Phys. Rev. E}\ }\textbf {\bibinfo {volume}
  {77}},\ \bibinfo {pages} {041304} (\bibinfo {year} {2008})}\BibitemShut
  {NoStop}%
\bibitem [{\citenamefont {Frenkel}\ \emph {et~al.}(2009)\citenamefont
  {Frenkel}, \citenamefont {Blumenfeld}, \citenamefont {King},\ and\
  \citenamefont {Blunt}}]{Fretal09}%
  \BibitemOpen
  \bibfield  {author} {\bibinfo {author} {\bibfnamefont {G.}~\bibnamefont
  {Frenkel}}, \bibinfo {author} {\bibfnamefont {R.}~\bibnamefont {Blumenfeld}},
  \bibinfo {author} {\bibfnamefont {P.}~\bibnamefont {King}}, \ and\ \bibinfo
  {author} {\bibfnamefont {M.}~\bibnamefont {Blunt}},\ }\href {\doibase
  10.1002/adem.200800090} {\bibfield  {journal} {\bibinfo  {journal} {Advanced
  engineering materials}\ }\textbf {\bibinfo {volume} {11}},\ \bibinfo {pages}
  {169} (\bibinfo {year} {2009})}\BibitemShut {NoStop}%
\bibitem [{\citenamefont {Hihinashvili}\ and\ \citenamefont
  {Blumenfeld}(2012)}]{HiBl12}%
  \BibitemOpen
  \bibfield  {author} {\bibinfo {author} {\bibfnamefont {R.}~\bibnamefont
  {Hihinashvili}}\ and\ \bibinfo {author} {\bibfnamefont {R.}~\bibnamefont
  {Blumenfeld}},\ }\href@noop {} {\bibfield  {journal} {\bibinfo  {journal}
  {Granular Matter}\ }\textbf {\bibinfo {volume} {14}},\ \bibinfo {pages} {277}
  (\bibinfo {year} {2012})}\BibitemShut {NoStop}%
\bibitem [{\citenamefont {Oda}(1982)}]{Oda1982}%
  \BibitemOpen
  \bibfield  {author} {\bibinfo {author} {\bibfnamefont {M.}~\bibnamefont
  {Oda}},\ }\href@noop {} {\bibfield  {journal} {\bibinfo  {journal} {Soils and
  Foundations}\ }\textbf {\bibinfo {volume} {22(4)}},\ \bibinfo {pages} {96}
  (\bibinfo {year} {1982})}\BibitemShut {NoStop}%
\bibitem [{\citenamefont {Satake}(1993)}]{Satake1993}%
  \BibitemOpen
  \bibfield  {author} {\bibinfo {author} {\bibfnamefont {M.}~\bibnamefont
  {Satake}},\ }\href@noop {} {\bibfield  {journal} {\bibinfo  {journal}
  {Mechanics of materials}\ }\textbf {\bibinfo {volume} {16(1)}},\ \bibinfo
  {pages} {65} (\bibinfo {year} {1993})}\BibitemShut {NoStop}%
\bibitem [{\citenamefont {Vogel}\ and\ \citenamefont
  {Roth}(2003)}]{Vogel_Roth_2003}%
  \BibitemOpen
  \bibfield  {author} {\bibinfo {author} {\bibfnamefont {H.-J.}\ \bibnamefont
  {Vogel}}\ and\ \bibinfo {author} {\bibfnamefont {K.}~\bibnamefont {Roth}},\
  }\href@noop {} {\bibfield  {journal} {\bibinfo  {journal} {Journal of
  Hydrology}\ }\textbf {\bibinfo {volume} {272}},\ \bibinfo {pages} {95}
  (\bibinfo {year} {2003})}\BibitemShut {NoStop}%
\bibitem [{\citenamefont {Cheng}\ \emph {et~al.}(1999)\citenamefont {Cheng},
  \citenamefont {Yu},\ and\ \citenamefont {Zulli}}]{Cheng_etal_1999}%
  \BibitemOpen
  \bibfield  {author} {\bibinfo {author} {\bibfnamefont {G.}~\bibnamefont
  {Cheng}}, \bibinfo {author} {\bibfnamefont {A.}~\bibnamefont {Yu}}, \ and\
  \bibinfo {author} {\bibfnamefont {P.}~\bibnamefont {Zulli}},\ }\href@noop {}
  {\bibfield  {journal} {\bibinfo  {journal} {Chem. Eng. Sci.}\ }\textbf
  {\bibinfo {volume} {54}},\ \bibinfo {pages} {4199} (\bibinfo {year}
  {1999})}\BibitemShut {NoStop}%
\bibitem [{\citenamefont {Paillusson}\ and\ \citenamefont
  {Frenkel}(2012)}]{PaFr12}%
  \BibitemOpen
  \bibfield  {author} {\bibinfo {author} {\bibfnamefont {F.}~\bibnamefont
  {Paillusson}}\ and\ \bibinfo {author} {\bibfnamefont {D.}~\bibnamefont
  {Frenkel}},\ }\href@noop {} {\bibfield  {journal} {\bibinfo  {journal} {Phys.
  Rev. Lett.}\ }\textbf {\bibinfo {volume} {109}},\ \bibinfo {pages} {208001}
  (\bibinfo {year} {2012})}\BibitemShut {NoStop}%
\bibitem [{\citenamefont {Blumenfeld}\ \emph {et~al.}(2015)\citenamefont
  {Blumenfeld}, \citenamefont {Edwards},\ and\ \citenamefont
  {Walley}}]{Bletal15}%
  \BibitemOpen
  \bibfield  {author} {\bibinfo {author} {\bibfnamefont {R.}~\bibnamefont
  {Blumenfeld}}, \bibinfo {author} {\bibfnamefont {S.~F.}\ \bibnamefont
  {Edwards}}, \ and\ \bibinfo {author} {\bibfnamefont {S.~M.}\ \bibnamefont
  {Walley}},\ }in\ \href@noop {} {\emph {\bibinfo {booktitle} {The Oxford
  Handbook of Soft Condensed Matter}}},\ Vol.\ \bibinfo {volume} {ISBN-13:
  978-0-19-966792-5},\ \bibinfo {editor} {edited by\ \bibinfo {editor}
  {\bibfnamefont {E.}~\bibnamefont {Terentjev}}\ and\ \bibinfo {editor}
  {\bibfnamefont {D.}~\bibnamefont {Weitz}}}\ (\bibinfo  {publisher} {Oxford
  University Press},\ \bibinfo {address} {Oxford, UK},\ \bibinfo {year}
  {2015})\BibitemShut {NoStop}%
\bibitem [{\citenamefont {Paillusson}(2015)}]{Pa15}%
  \BibitemOpen
  \bibfield  {author} {\bibinfo {author} {\bibfnamefont {F.}~\bibnamefont
  {Paillusson}},\ }\href@noop {} {\bibfield  {journal} {\bibinfo  {journal}
  {Phys. Rev. E.}\ }\textbf {\bibinfo {volume} {91}},\ \bibinfo {pages}
  {012204} (\bibinfo {year} {2015})}\BibitemShut {NoStop}%
\bibitem [{\citenamefont {Frenkel}\ \emph {et~al.}(2013)\citenamefont
  {Frenkel}, \citenamefont {Asenjo},\ and\ \citenamefont
  {Paillusson}}]{Fretal13}%
  \BibitemOpen
  \bibfield  {author} {\bibinfo {author} {\bibfnamefont {D.}~\bibnamefont
  {Frenkel}}, \bibinfo {author} {\bibfnamefont {D.}~\bibnamefont {Asenjo}}, \
  and\ \bibinfo {author} {\bibfnamefont {F.}~\bibnamefont {Paillusson}},\
  }\href@noop {} {\bibfield  {journal} {\bibinfo  {journal} {Mol. Phys.}\
  }\textbf {\bibinfo {volume} {111}},\ \bibinfo {pages} {3641} (\bibinfo {year}
  {2013})}\BibitemShut {NoStop}%
\bibitem [{\citenamefont {Bi}\ \emph {et~al.}(2015)\citenamefont {Bi},
  \citenamefont {Henkes}, \citenamefont {Daniels},\ and\ \citenamefont
  {Chakraborty}}]{Bietal15}%
  \BibitemOpen
  \bibfield  {author} {\bibinfo {author} {\bibfnamefont {D.}~\bibnamefont
  {Bi}}, \bibinfo {author} {\bibfnamefont {S.}~\bibnamefont {Henkes}}, \bibinfo
  {author} {\bibfnamefont {K.~E.}\ \bibnamefont {Daniels}}, \ and\ \bibinfo
  {author} {\bibfnamefont {B.}~\bibnamefont {Chakraborty}},\ }\href@noop {}
  {\bibfield  {journal} {\bibinfo  {journal} {Annual Rev. Cond. Matt. Phys.}\
  }\textbf {\bibinfo {volume} {6}},\ \bibinfo {pages} {63} (\bibinfo {year}
  {2015})}\BibitemShut {NoStop}%
\bibitem [{\citenamefont {Edwards}\ and\ \citenamefont
  {Blumenfeld}(2005)}]{BlEd05}%
  \BibitemOpen
  \bibfield  {author} {\bibinfo {author} {\bibfnamefont {S.~F.}\ \bibnamefont
  {Edwards}}\ and\ \bibinfo {author} {\bibfnamefont {R.}~\bibnamefont
  {Blumenfeld}},\ }in\ \href@noop {} {\emph {\bibinfo {booktitle} {Powders and
  Grains, Stuttgart}}},\ \bibinfo {editor} {edited by\ \bibinfo {editor}
  {\bibfnamefont {R.}~\bibnamefont {Garcia-Rojo}}, \bibinfo {editor}
  {\bibfnamefont {H.~J.}\ \bibnamefont {Herrmann}}, \ and\ \bibinfo {editor}
  {\bibfnamefont {S.}~\bibnamefont {McNamara}}}\ (\bibinfo  {publisher}
  {Balkema},\ \bibinfo {address} {Leiden, Netherlands},\ \bibinfo {year}
  {2005})\ pp.\ \bibinfo {pages} {3--5}\BibitemShut {NoStop}%
\bibitem [{\citenamefont {Henkes}\ \emph {et~al.}(2007)\citenamefont {Henkes},
  \citenamefont {O'Hern},\ and\ \citenamefont {Chakraborty}}]{Heetal07}%
  \BibitemOpen
  \bibfield  {author} {\bibinfo {author} {\bibfnamefont {S.}~\bibnamefont
  {Henkes}}, \bibinfo {author} {\bibfnamefont {C.~S.}\ \bibnamefont {O'Hern}},
  \ and\ \bibinfo {author} {\bibfnamefont {B.}~\bibnamefont {Chakraborty}},\
  }\href@noop {} {\bibfield  {journal} {\bibinfo  {journal} {Phys. Rev. Lett.}\
  }\textbf {\bibinfo {volume} {99}},\ \bibinfo {pages} {038002} (\bibinfo
  {year} {2007})}\BibitemShut {NoStop}%
\bibitem [{\citenamefont {Edwards}\ and\ \citenamefont
  {Oakeshott}(1989{\natexlab{a}})}]{EdOa89a}%
  \BibitemOpen
  \bibfield  {author} {\bibinfo {author} {\bibfnamefont {S.}~\bibnamefont
  {Edwards}}\ and\ \bibinfo {author} {\bibfnamefont {R.}~\bibnamefont
  {Oakeshott}},\ }\href@noop {} {\bibfield  {journal} {\bibinfo  {journal}
  {Physica D}\ }\textbf {\bibinfo {volume} {38}},\ \bibinfo {pages} {88}
  (\bibinfo {year} {1989}{\natexlab{a}})}\BibitemShut {NoStop}%
\bibitem [{\citenamefont {Matheson}(1971)}]{Ma71}%
  \BibitemOpen
  \bibfield  {author} {\bibinfo {author} {\bibfnamefont {J.~A.~L.}\
  \bibnamefont {Matheson}},\ }\href@noop {} {\emph {\bibinfo {title}
  {Hyperstatic structures: an introduction to the theory of statically
  indeterminate structures}}}\ (\bibinfo  {publisher} {Butterworths},\ \bibinfo
  {year} {1971})\BibitemShut {NoStop}%
\bibitem [{\citenamefont {Blumenfeld}\ \emph {et~al.}(2016)\citenamefont
  {Blumenfeld}, \citenamefont {Amitai}, \citenamefont {Jordan},\ and\
  \citenamefont {Hihinashvili}}]{Bletal16}%
  \BibitemOpen
  \bibfield  {author} {\bibinfo {author} {\bibfnamefont {R.}~\bibnamefont
  {Blumenfeld}}, \bibinfo {author} {\bibfnamefont {S.}~\bibnamefont {Amitai}},
  \bibinfo {author} {\bibfnamefont {J.~F.}\ \bibnamefont {Jordan}}, \ and\
  \bibinfo {author} {\bibfnamefont {R.}~\bibnamefont {Hihinashvili}},\
  }\href@noop {} {\bibfield  {journal} {\bibinfo  {journal} {Phys. Rev. Lett.}\
  }\textbf {\bibinfo {volume} {116}},\ \bibinfo {pages} {148001} (\bibinfo
  {year} {2016})}\BibitemShut {NoStop}%
\bibitem [{\citenamefont {Blumenfeld}\ and\ \citenamefont
  {Edwards}(2009)}]{BlEd09}%
  \BibitemOpen
  \bibfield  {author} {\bibinfo {author} {\bibfnamefont {R.}~\bibnamefont
  {Blumenfeld}}\ and\ \bibinfo {author} {\bibfnamefont {S.}~\bibnamefont
  {Edwards}},\ }\href@noop {} {\bibfield  {journal} {\bibinfo  {journal} {J.
  Phys. Chem. B}\ }\textbf {\bibinfo {volume} {113}},\ \bibinfo {pages} {3981}
  (\bibinfo {year} {2009})}\BibitemShut {NoStop}%
\bibitem [{\citenamefont {Cundall}\ and\ \citenamefont
  {Strack}(1979)}]{Cundall1979}%
  \BibitemOpen
  \bibfield  {author} {\bibinfo {author} {\bibfnamefont {P.~A.}\ \bibnamefont
  {Cundall}}\ and\ \bibinfo {author} {\bibfnamefont {O.~D.~L.}\ \bibnamefont
  {Strack}},\ }\href@noop {} {\bibfield  {journal} {\bibinfo  {journal}
  {Geotechnique}\ }\textbf {\bibinfo {volume} {29}},\ \bibinfo {pages} {47}
  (\bibinfo {year} {1979})}\BibitemShut {NoStop}%
\bibitem [{\citenamefont {Matsushima}\ and\ \citenamefont
  {Chang}(2011)}]{Mat-Chang-2011}%
  \BibitemOpen
  \bibfield  {author} {\bibinfo {author} {\bibfnamefont {T.}~\bibnamefont
  {Matsushima}}\ and\ \bibinfo {author} {\bibfnamefont {C.~S.}\ \bibnamefont
  {Chang}},\ }\href@noop {} {\bibfield  {journal} {\bibinfo  {journal}
  {Granular matter}\ }\textbf {\bibinfo {volume} {13}},\ \bibinfo {pages} {269}
  (\bibinfo {year} {2011})}\BibitemShut {NoStop}%
\bibitem [{\citenamefont {Hakuno}(1997)}]{Hakuno97}%
  \BibitemOpen
  \bibfield  {author} {\bibinfo {author} {\bibfnamefont {M.}~\bibnamefont
  {Hakuno}},\ }\href@noop {} {\emph {\bibinfo {title} {Numerical simulation for
  Failure}}}\ (\bibinfo  {publisher} {Morikita Press},\ \bibinfo {year}
  {1997})\BibitemShut {NoStop}%
\bibitem [{\citenamefont {Calvetti}(2008)}]{Calvetti08}%
  \BibitemOpen
  \bibfield  {author} {\bibinfo {author} {\bibfnamefont {F.~G.}\ \bibnamefont
  {Calvetti}},\ }\href@noop {} {\bibfield  {journal} {\bibinfo  {journal}
  {European Journal of Environmental and Civil Engineering}\ }\textbf {\bibinfo
  {volume} {12(7-8)}},\ \bibinfo {pages} {951} (\bibinfo {year}
  {2008})}\BibitemShut {NoStop}%
\bibitem [{\citenamefont {Mitchell}\ and\ \citenamefont
  {Soga}(2005)}]{Mitchell-Soga-2005}%
  \BibitemOpen
  \bibfield  {author} {\bibinfo {author} {\bibfnamefont {J.~K.}\ \bibnamefont
  {Mitchell}}\ and\ \bibinfo {author} {\bibfnamefont {K.}~\bibnamefont
  {Soga}},\ }\href@noop {} {\emph {\bibinfo {title} {Fundamentals of Soil
  Behavior}}},\ \bibinfo {edition} {3rd}\ ed.\ (\bibinfo  {publisher} {John
  Wiley \& Sons},\ \bibinfo {year} {2005})\BibitemShut {NoStop}%
\bibitem [{\citenamefont {Matsushima}\ and\ \citenamefont
  {Blumenfeld}(2013)}]{MaBl13}%
  \BibitemOpen
  \bibfield  {author} {\bibinfo {author} {\bibfnamefont {T.}~\bibnamefont
  {Matsushima}}\ and\ \bibinfo {author} {\bibfnamefont {R.}~\bibnamefont
  {Blumenfeld}},\ }\href@noop {} {\bibfield  {journal} {\bibinfo  {journal}
  {AIP Conference Proceedings (Powders and Grains 2013)}\ }\textbf {\bibinfo
  {volume} {1542(1)}},\ \bibinfo {pages} {325} (\bibinfo {year}
  {2013})}\BibitemShut {NoStop}%
\bibitem [{\citenamefont {Silbert}(2010)}]{silbert2010}%
  \BibitemOpen
  \bibfield  {author} {\bibinfo {author} {\bibfnamefont {L.~E.}\ \bibnamefont
  {Silbert}},\ }\href@noop {} {\bibfield  {journal} {\bibinfo  {journal} {Soft
  Matter}\ }\textbf {\bibinfo {volume} {6}},\ \bibinfo {pages} {2918} (\bibinfo
  {year} {2010})}\BibitemShut {NoStop}%
\bibitem [{\citenamefont {Abate}\ and\ \citenamefont
  {Durian}(2006)}]{abate2006approach}%
  \BibitemOpen
  \bibfield  {author} {\bibinfo {author} {\bibfnamefont {A.~R.}\ \bibnamefont
  {Abate}}\ and\ \bibinfo {author} {\bibfnamefont {D.~J.}\ \bibnamefont
  {Durian}},\ }\href@noop {} {\bibfield  {journal} {\bibinfo  {journal}
  {Physical Review E}\ }\textbf {\bibinfo {volume} {74}},\ \bibinfo {pages}
  {031308} (\bibinfo {year} {2006})}\BibitemShut {NoStop}%
\bibitem [{\citenamefont {Blumenfeld}\ \emph {et~al.}()\citenamefont
  {Blumenfeld}, \citenamefont {Toikka},\ and\ \citenamefont
  {Matsushima}}]{BlToMaSoon}%
  \BibitemOpen
  \bibfield  {author} {\bibinfo {author} {\bibfnamefont {R.}~\bibnamefont
  {Blumenfeld}}, \bibinfo {author} {\bibfnamefont {L.}~\bibnamefont {Toikka}},
  \ and\ \bibinfo {author} {\bibfnamefont {T.}~\bibnamefont {Matsushima}},\
  }\href@noop {} {\bibinfo  {journal} {In preparation}\ }\BibitemShut {NoStop}%
\bibitem [{\citenamefont {Edwards}\ and\ \citenamefont
  {Oakeshott}(1989{\natexlab{b}})}]{EdOa89b}%
  \BibitemOpen
\bibfield  {journal} {  }\bibfield  {author} {\bibinfo {author} {\bibfnamefont
  {S.}~\bibnamefont {Edwards}}\ and\ \bibinfo {author} {\bibfnamefont
  {R.}~\bibnamefont {Oakeshott}},\ }\href@noop {} {\bibfield  {journal}
  {\bibinfo  {journal} {Physica A}\ }\textbf {\bibinfo {volume} {157}},\
  \bibinfo {pages} {1080} (\bibinfo {year} {1989}{\natexlab{b}})}\BibitemShut
  {NoStop}%
\bibitem [{\citenamefont {Xu}\ \emph {et~al.}(2005)\citenamefont {Xu},
  \citenamefont {Blawzdziewicz},\ and\ \citenamefont {OfHern}}]{xu2005random}%
  \BibitemOpen
  \bibfield  {author} {\bibinfo {author} {\bibfnamefont {N.}~\bibnamefont
  {Xu}}, \bibinfo {author} {\bibfnamefont {J.}~\bibnamefont {Blawzdziewicz}}, \
  and\ \bibinfo {author} {\bibfnamefont {C.~S.}\ \bibnamefont {OfHern}},\
  }\href@noop {} {\bibfield  {journal} {\bibinfo  {journal} {Physical Review
  E}\ }\textbf {\bibinfo {volume} {71}},\ \bibinfo {pages} {061306} (\bibinfo
  {year} {2005})}\BibitemShut {NoStop}%
\bibitem [{\citenamefont {Gao}\ \emph {et~al.}(2006)\citenamefont {Gao},
  \citenamefont {B{\l}awzdziewicz},\ and\ \citenamefont
  {OfHern}}]{gao2006frequency}%
  \BibitemOpen
  \bibfield  {author} {\bibinfo {author} {\bibfnamefont {G.-J.}\ \bibnamefont
  {Gao}}, \bibinfo {author} {\bibfnamefont {J.}~\bibnamefont
  {B{\l}awzdziewicz}}, \ and\ \bibinfo {author} {\bibfnamefont {C.~S.}\
  \bibnamefont {OfHern}},\ }\href@noop {} {\bibfield  {journal} {\bibinfo
  {journal} {Physical Review E}\ }\textbf {\bibinfo {volume} {74}},\ \bibinfo
  {pages} {061304} (\bibinfo {year} {2006})}\BibitemShut {NoStop}%
\bibitem [{\citenamefont {Gao}\ \emph {et~al.}(2007)\citenamefont {Gao},
  \citenamefont {Blawzdziewicz},\ and\ \citenamefont
  {O'Hern}}]{gao2007enumeration}%
  \BibitemOpen
  \bibfield  {author} {\bibinfo {author} {\bibfnamefont {G.-J.}\ \bibnamefont
  {Gao}}, \bibinfo {author} {\bibfnamefont {J.}~\bibnamefont {Blawzdziewicz}},
  \ and\ \bibinfo {author} {\bibfnamefont {C.~S.}\ \bibnamefont {O'Hern}},\
  }\href@noop {} {\bibfield  {journal} {\bibinfo  {journal} {Philosophical
  Magazine}\ }\textbf {\bibinfo {volume} {87}},\ \bibinfo {pages} {425}
  (\bibinfo {year} {2007})}\BibitemShut {NoStop}%
\bibitem [{\citenamefont {Gao}\ \emph {et~al.}(2009{\natexlab{a}})\citenamefont
  {Gao}, \citenamefont {Blawzdziewicz},\ and\ \citenamefont
  {OfHern}}]{gao2009geometrical}%
  \BibitemOpen
  \bibfield  {author} {\bibinfo {author} {\bibfnamefont {G.-J.}\ \bibnamefont
  {Gao}}, \bibinfo {author} {\bibfnamefont {J.}~\bibnamefont {Blawzdziewicz}},
  \ and\ \bibinfo {author} {\bibfnamefont {C.~S.}\ \bibnamefont {OfHern}},\
  }\href@noop {} {\bibfield  {journal} {\bibinfo  {journal} {Physical Review
  E}\ }\textbf {\bibinfo {volume} {80}},\ \bibinfo {pages} {061303} (\bibinfo
  {year} {2009}{\natexlab{a}})}\BibitemShut {NoStop}%
\bibitem [{\citenamefont {Gao}\ \emph {et~al.}(2009{\natexlab{b}})\citenamefont
  {Gao}, \citenamefont {Blawzdziewicz}, \citenamefont {OfHern},\ and\
  \citenamefont {Shattuck}}]{gao2009experimental}%
  \BibitemOpen
  \bibfield  {author} {\bibinfo {author} {\bibfnamefont {G.-J.}\ \bibnamefont
  {Gao}}, \bibinfo {author} {\bibfnamefont {J.}~\bibnamefont {Blawzdziewicz}},
  \bibinfo {author} {\bibfnamefont {C.~S.}\ \bibnamefont {OfHern}}, \ and\
  \bibinfo {author} {\bibfnamefont {M.}~\bibnamefont {Shattuck}},\ }\href@noop
  {} {\bibfield  {journal} {\bibinfo  {journal} {Physical Review E}\ }\textbf
  {\bibinfo {volume} {80}},\ \bibinfo {pages} {061304} (\bibinfo {year}
  {2009}{\natexlab{b}})}\BibitemShut {NoStop}%
\bibitem [{\citenamefont {Smith}\ \emph {et~al.}(1929)\citenamefont {Smith},
  \citenamefont {Foote},\ and\ \citenamefont {Busang}}]{smith1929packing}%
  \BibitemOpen
  \bibfield  {author} {\bibinfo {author} {\bibfnamefont {W.}~\bibnamefont
  {Smith}}, \bibinfo {author} {\bibfnamefont {P.~D.}\ \bibnamefont {Foote}}, \
  and\ \bibinfo {author} {\bibfnamefont {P.}~\bibnamefont {Busang}},\
  }\href@noop {} {\bibfield  {journal} {\bibinfo  {journal} {Physical Review}\
  }\textbf {\bibinfo {volume} {34}},\ \bibinfo {pages} {1271} (\bibinfo {year}
  {1929})}\BibitemShut {NoStop}%
\bibitem [{\citenamefont {Katagiri}\ \emph {et~al.}(2014)\citenamefont
  {Katagiri}, \citenamefont {Matsushima},\ and\ \citenamefont
  {Yamada}}]{katagiri2014variations}%
  \BibitemOpen
  \bibfield  {author} {\bibinfo {author} {\bibfnamefont {J.}~\bibnamefont
  {Katagiri}}, \bibinfo {author} {\bibfnamefont {T.}~\bibnamefont
  {Matsushima}}, \ and\ \bibinfo {author} {\bibfnamefont {Y.}~\bibnamefont
  {Yamada}},\ }\href@noop {} {\bibfield  {journal} {\bibinfo  {journal}
  {Granular Matter}\ }\textbf {\bibinfo {volume} {16}},\ \bibinfo {pages} {891}
  (\bibinfo {year} {2014})}\BibitemShut {NoStop}%
\bibitem [{\citenamefont {Meyer}\ \emph {et~al.}(2010)\citenamefont {Meyer},
  \citenamefont {Song}, \citenamefont {Jin}, \citenamefont {Wang},\ and\
  \citenamefont {Makse}}]{Meetal10}%
  \BibitemOpen
  \bibfield  {author} {\bibinfo {author} {\bibfnamefont {S.}~\bibnamefont
  {Meyer}}, \bibinfo {author} {\bibfnamefont {C.}~\bibnamefont {Song}},
  \bibinfo {author} {\bibfnamefont {Y.}~\bibnamefont {Jin}}, \bibinfo {author}
  {\bibfnamefont {K.}~\bibnamefont {Wang}}, \ and\ \bibinfo {author}
  {\bibfnamefont {H.}~\bibnamefont {Makse}},\ }\href@noop {} {\bibfield
  {journal} {\bibinfo  {journal} {Physica A}\ }\textbf {\bibinfo {volume}
  {389}},\ \bibinfo {pages} {5137} (\bibinfo {year} {2010})}\BibitemShut
  {NoStop}%
\end{thebibliography}%

\end{document}